\input harvmac   
\noblackbox   

\input labeldefs.tmp


\def\p{\partial}

\def\CN{{\cal N}}

\def\CL{{\cal L}}

\def\CS{{\cal S}}

\def\CS{{\cal S }}

\def\G{\Gamma}    


\font\manual=manfnt     
\def\dbend{\lower3.5pt\hbox{\manual\char127}}

\def\IZ{\relax\ifmmode\mathchoice    
{\hbox{\cmss Z\kern-.4em Z}}{\hbox{\cmss Z\kern-.4em Z}}    
{\lower.9pt\hbox{\cmsss Z\kern-.4em Z}} {\lower1.2pt\hbox{\cmsss    
Z\kern-.4em Z}}\else{\cmss Z\kern-.4em Z}\fi}    
\def\half {{1\over 2}}

\def\p{\partial}    
    
\def\bar{\overline}    
\def\CS{{\cal S}}    
\def\CN{{\cal N}}    
    
\def\rt2{\sqrt{2}}    
\def\irt2{{1\over\sqrt{2}}}

\def\t{\tilde}    
\def\ndt{\noindent}    
\def\s{\sigma}

\font\cmss=cmss10    
\font\cmsss=cmss10 at 7pt    
\def\IL{\relax{\rm I\kern-.18em L}}    
\def\IH{\relax{\rm I\kern-.18em H}}    
\def\IR{\relax{\rm I\kern-.18em R}}    
\def\inbar{\vrule height1.5ex width.4pt depth0pt}    
\def\IC{\relax\hbox{$\inbar\kern-.3em{\rm C}$}}    
\def\rlx{\relax\leavevmode}    
\def\ZZ{\rlx\leavevmode\ifmmode\mathchoice{\hbox{\cmss Z\kern-.4em Z}}    
 {\hbox{\cmss Z\kern-.4em Z}}{\lower.9pt\hbox{\cmsss Z\kern-.36em Z}}    
 {\lower1.2pt\hbox{\cmsss Z\kern-.36em Z}}\else{\cmss Z\kern-.4em    
 Z}\fi}     
\def\IZ{\relax\ifmmode\mathchoice    
{\hbox{\cmss Z\kern-.4em Z}}{\hbox{\cmss Z\kern-.4em Z}}    
{\lower.9pt\hbox{\cmsss Z\kern-.4em Z}}    
{\lower1.2pt\hbox{\cmsss Z\kern-.4em Z}}\else{\cmss Z\kern-.4em    
Z}\fi}    
    

\def\G{\Gamma}

\font\manual=manfnt     
\def\dbend{\lower3.5pt\hbox{\manual\char127}}

\def\IZ{\relax\ifmmode\mathchoice    
{\hbox{\cmss Z\kern-.4em Z}}{\hbox{\cmss Z\kern-.4em Z}}    
{\lower.9pt\hbox{\cmsss Z\kern-.4em Z}} {\lower1.2pt\hbox{\cmsss    
Z\kern-.4em Z}}\else{\cmss Z\kern-.4em Z}\fi}    
\def\half {{1\over 2}}

\def\bar{\overline}    

\def\rt2{\sqrt{2}}    
\def\irt2{{1\over\sqrt{2}}}

\def\t{\tilde}    
\def\T{\widetilde}
\def\ndt{\noindent}    
\def\s{\sigma}

\def\bra#1{{\langle}#1|}
\def\ket#1{|#1\rangle}
\def\bbra#1{{\langle\langle}#1|}
\def\kket#1{|#1\rangle\rangle}

\let\includefigures=\iftrue
\let\useblackboard=\iftrue
\newfam\black

\includefigures
\message{If you do not have epsf.tex (to include figures),}
\message{change the option at the top of the tex file.}
\input epsf
\def\figin{\epsfcheck\figin}\def\figins{\epsfcheck\figins}
\def\epsfcheck{\ifx\epsfbox\UnDeFiNeD
\message{(NO epsf.tex, FIGURES WILL BE IGNORED)}
\gdef\figin##1{\vskip2in}\gdef\figins##1{\hskip.5in}
\else\message{(FIGURES WILL BE INCLUDED)}%
\gdef\figin##1{##1}\gdef\figins##1{##1}\fi}
\def\DefWarn#1{}
\def\figinsert{\goodbreak\midinsert}
\def\ifig#1#2#3{\DefWarn#1\xdef#1{fig.~\the\figno}
\writedef{#1\leftbracket fig.\noexpand~\the\figno}%
\figinsert\figin{\centerline{#3}}\medskip\centerline{\vbox{
\baselineskip12pt\advance\hsize by -1truein
\noindent\footnotefont{\bf Fig.~\the\figno:} #2}}
\bigskip\endinsert\global\advance\figno by1}
\else
\def\ifig#1#2#3{\xdef#1{fig.~\the\figno}
\writedef{#1\leftbracket fig.\noexpand~\the\figno}%
\global\advance\figno by1}
\fi

\def\doublefig#1#2#3#4{\DefWarn#1\xdef#1{fig.~\the\figno}
\writedef{#1\leftbracket fig.\noexpand~\the\figno}%
\figinsert\figin{\centerline{#3\hskip1.0cm#4}}\medskip\centerline{\vbox{
\baselineskip12pt\advance\hsize by -1truein
\noindent\footnotefont{\bf Fig.~\the\figno:} #2}}
\bigskip\endinsert\global\advance\figno by1}


\writedefs


\lref\mcgreevy{
  J.~McGreevy and H.~Verlinde,
  ``Strings from tachyons: The c = 1 matrix reloaded,''
  JHEP {\bf 0312}, 054 (2003)
  [arXiv:hep-th/0304224] \semi
  I.~R.~Klebanov, J.~Maldacena and N.~Seiberg,
  ``D-brane decay in two-dimensional string theory,''
  JHEP {\bf 0307}, 045 (2003)
  [arXiv:hep-th/0305159] \semi
  M.~R.~Douglas, I.~R.~Klebanov, D.~Kutasov, J.~Maldacena, E.~Martinec and N.~Seiberg,
  ``A new hat for the c = 1 matrix model,''
  arXiv:hep-th/0307195 \semi
  T.~Takayanagi and N.~Toumbas,
  ``A matrix model dual of type 0B string theory in two dimensions,''
  JHEP {\bf 0307}, 064 (2003)
  [arXiv:hep-th/0307083].
}

\lref\McGreevyEP{
  J.~McGreevy, J.~Teschner and H.~Verlinde,
  ``Classical and quantum D-branes in 2D string theory,''
  JHEP {\bf 0401}, 039 (2004)
  [arXiv:hep-th/0305194].
}

\lref\DiVecchiaNE{
  P.~Di Vecchia and A.~Liccardo,
  ``Gauge theories from D branes,''
  arXiv:hep-th/0307104.
}

\lref\klebstras{
  I.~R.~Klebanov and M.~J.~Strassler,
  ``Supergravity and a confining gauge theory: Duality cascades and
  chiSB-resolution of naked singularities,''
  JHEP {\bf 0008}, 052 (2000)
  [arXiv:hep-th/0007191].
}

\lref\polstras{
  J.~Polchinski and M.~J.~Strassler,
  ``The string dual of a confining four-dimensional gauge theory,''
  arXiv:hep-th/0003136.
}

\lref\maldanunez{
  J.~M.~Maldacena and C.~Nunez,
  ``Towards the large N limit of pure N = 1 super Yang Mills,''
  Phys.\ Rev.\ Lett.\  {\bf 86}, 588 (2001)
  [arXiv:hep-th/0008001].
}

\lref\maldalunin{
  O.~Lunin and J.~Maldacena,
  ``Deforming field theories with U(1) x U(1) global symmetry and their gravity
  duals,''
  arXiv:hep-th/0502086.
}

\lref\HananyIE{
  A.~Hanany and E.~Witten,
  ``Type IIB superstrings, BPS monopoles, and three-dimensional gauge dynamics,''
  Nucl.\ Phys.\ B {\bf 492}, 152 (1997)
  [arXiv:hep-th/9611230].
}

\lref\PolchinskiMT{
  J.~Polchinski,
  ``Dirichlet-Branes and Ramond-Ramond Charges,''
  Phys.\ Rev.\ Lett.\  {\bf 75}, 4724 (1995)
  [arXiv:hep-th/9510017].
}

\lref\MaldacenaRE{
  J.~M.~Maldacena,
  ``The large N limit of superconformal field theories and supergravity,''
  Adv.\ Theor.\ Math.\ Phys.\  {\bf 2}, 231 (1998)
  [Int.\ J.\ Theor.\ Phys.\  {\bf 38}, 1113 (1999)]
  [arXiv:hep-th/9711200].
}

\lref\EguchiTC{
T.~Eguchi and Y.~Sugawara,
``Modular invariance in superstring on Calabi-Yau n-fold with A-D-E
singularity,''
Nucl.\ Phys.\ B {\bf 577}, 3 (2000)
[arXiv:hep-th/0002100].
}

\lref\MurthyES{
S.~Murthy,
``Notes on non-critical superstrings in various dimensions,''
JHEP {\bf 0311}, 056 (2003)
[arXiv:hep-th/0305197].
}

\lref\KlebanovYA{
I.~R.~Klebanov and J.~M.~Maldacena,
``Superconformal gauge theories and non-critical superstrings,''
arXiv:hep-th/0409133.
}

\lref\KS{S. Kuperstein and J. Sonnenschein,
  ``Non-critical supergravity ($d > 1$) and holography,''
  JHEP {\bf 0407}, 049 (2004) [arXiv:hep-th/0403254].}

\lref\KazamaQP{
  Y.~Kazama and H.~Suzuki,
  ``New N=2 Superconformal Field Theories And Superstring Compactification,''
  Nucl.\ Phys.\ B {\bf 321}, 232 (1989).
}

\lref\ZamolodchikovAH{
A.~B.~Zamolodchikov and A.~B.~Zamolodchikov,
``Liouville field theory on a pseudosphere,''
arXiv:hep-th/0101152.
}

\lref\ElitzurPQ{
S.~Elitzur, A.~Giveon, D.~Kutasov, E.~Rabinovici and G.~Sarkissian,
``D-branes in the background of NS fivebranes,''
JHEP {\bf 0008}, 046 (2000)
[arXiv:hep-th/0005052].
}

\lref\BoucherBH{
W.~Boucher, D.~Friedan and A.~Kent,
``Determinant Formulae And Unitarity For The N=2 Superconformal Algebras In
Two-Dimensions Or Exact Results On String Compactification,''
Phys.\ Lett.\ B {\bf 172}, 316 (1986).
}

\lref\NamHU{
S.~k.~Nam,
``Superconformal And Super Kac-Moody Invariant Quantum Field Theories In
Two-Dimensions,''
Phys.\ Lett.\ B {\bf 187}, 340 (1987).
}

\lref\KiritsisRV{
E.~Kiritsis,
``Character Formulae And The Structure Of The Representations Of The N=1, N=2
Superconformal Algebras,''
Int.\ J.\ Mod.\ Phys.\ A {\bf 3}, 1871 (1988).
}

\lref\AharonyXN{
O.~Aharony, A.~Giveon and D.~Kutasov,
``LSZ in LST,''
Nucl.\ Phys.\ B {\bf 691}, 3 (2004)
[arXiv:hep-th/0404016].
}

\lref\Giveon{
  A.~Giveon, A.~Konechny, A.~Pakman and A.~Sever,
  ``Type 0 strings in a 2-d black hole,''
  JHEP {\bf 0310}, 025 (2003)
  [arXiv:hep-th/0309056].
}

\lref\WittenYR{
E.~Witten,
``On string theory and black holes,''
Phys.\ Rev.\ D {\bf 44}, 314 (1991).
}

\lref\DijkgraafBA{
R.~Dijkgraaf, H.~Verlinde and E.~Verlinde,
``String propagation in a black hole geometry,''
Nucl.\ Phys.\ B {\bf 371}, 269 (1992).
}

\lref\HoriAX{
K.~Hori and A.~Kapustin,
``Duality of the fermionic 2d black hole and N = 2 Liouville theory as  mirror
symmetry,''
JHEP {\bf 0108}, 045 (2001)
[arXiv:hep-th/0104202].
}

\lref\AhnTT{
  C.~Ahn, M.~Stanishkov and M.~Yamamoto,
  Nucl.\ Phys.\ B {\bf 683}, 177 (2004)
  [arXiv:hep-th/0311169].
}

\lref\OoguriCK{
  H.~Ooguri, Y.~Oz and Z.~Yin,
  ``D-branes on Calabi-Yau spaces and their mirrors,''
  Nucl.\ Phys.\ B {\bf 477}, 407 (1996)
  [arXiv:hep-th/9606112].
}

\lref\KutasovPV{
D.~Kutasov,
``Some properties of (non)critical strings,''
arXiv:hep-th/9110041.
}

\lref\KutasovUA{
D.~Kutasov and N.~Seiberg,
``Noncritical Superstrings,''
Phys.\ Lett.\ B {\bf 251}, 67 (1990).
}

\lref\EguchiYI{
T.~Eguchi and Y.~Sugawara,
``SL(2,R)/U(1) supercoset and elliptic genera of non-compact Calabi-Yau
manifolds,''
JHEP {\bf 0405}, 014 (2004)
[arXiv:hep-th/0403193].
}

\lref\FotopoulosUT{
A.~Fotopoulos, V.~Niarchos and N.~Prezas,
``D-branes and extended characters in SL(2,R)/U(1),''
Nucl.\ Phys.\ B {\bf 710}, 309 (2005)
[arXiv:hep-th/0406017].
}

\lref\DiVecchiaRH{
P.~Di Vecchia and A.~Liccardo,
``D branes in string theory. I,''
NATO Adv.\ Study Inst.\ Ser.\ C.\ Math.\ Phys.\ Sci.\  {\bf 556}, 1 (2000)
[arXiv:hep-th/9912161].
}

\lref\DiVecchiaFX{
P.~Di Vecchia and A.~Liccardo,
``D-branes in string theory. II,''
arXiv:hep-th/9912275.
}

\lref\GaberdielJR{
M.~R.~Gaberdiel,
``Lectures on non-BPS Dirichlet branes,''
Class.\ Quant.\ Grav.\  {\bf 17}, 3483 (2000)
[arXiv:hep-th/0005029].
}

\lref\JegoTA{
  C.~Jego and J.~Troost,
  ``Notes on the Verlinde formula in non-rational conformal field theories,''
  arXiv:hep-th/0601085.
}

\lref\HananyTB{
  A.~Hanany and A.~Zaffaroni,
  ``On the realization of chiral four-dimensional gauge theories using
  branes,''
  JHEP {\bf 9805}, 001 (1998)
  [arXiv:hep-th/9801134].
}

\lref\FotopoulosUT{
  A.~Fotopoulos, V.~Niarchos and N.~Prezas,
  ``D-branes and extended characters in SL(2,R)/U(1),''
  Nucl.\ Phys.\ B {\bf 710}, 309 (2005)
  [arXiv:hep-th/0406017].
}

\lref\GiveonPX{
A.~Giveon and D.~Kutasov,
``Little string theory in a double scaling limit,''
JHEP {\bf 9910}, 034 (1999)
[arXiv:hep-th/9909110].
}

\lref\FZZ{
V.~A.~Fateev, A.~B.~Zamolodchikov and Al.~B.~Zamolodchikov, unpublished.
}

\lref\AharonyUB{
O.~Aharony, M.~Berkooz, D.~Kutasov and N.~Seiberg,
``Linear dilatons, NS5-branes and holography,''
JHEP {\bf 9810}, 004 (1998)
[arXiv:hep-th/9808149].
}

\lref\GiveonZM{
A.~Giveon, D.~Kutasov and O.~Pelc,
``Holography for non-critical superstrings,''
JHEP {\bf 9910}, 035 (1999)
[arXiv:hep-th/9907178].
}

\lref\HikidaXU{
  Y.~Hikida and Y.~Sugawara,
  ``Superstring vacua of 4-dimensional pp-waves with enhanced  supersymmetry,''
  JHEP {\bf 0210}, 067 (2002)
  [arXiv:hep-th/0207124].
}

\lref\RecknagelSB{
  A.~Recknagel and V.~Schomerus,
  ``D-branes in Gepner models,''
  Nucl.\ Phys.\ B {\bf 531}, 185 (1998)
  [arXiv:hep-th/9712186].
}

\lref\GutperleHB{
  M.~Gutperle and Y.~Satoh,
  ``D-branes in Gepner models and supersymmetry,''
  Nucl.\ Phys.\ B {\bf 543}, 73 (1999)
  [arXiv:hep-th/9808080].
}

\lref\WittenSC{
  E.~Witten,
  ``Solutions of four-dimensional field theories via M-theory,''
  Nucl.\ Phys.\ B {\bf 500}, 3 (1997)
  [arXiv:hep-th/9703166].
}

\lref\ElitzurPQ{
S.~Elitzur, A.~Giveon, D.~Kutasov, E.~Rabinovici and G.~Sarkissian,
``D-branes in the background of NS fivebranes,''
JHEP {\bf 0008}, 046 (2000)
[arXiv:hep-th/0005052].
}

\lref\FateevIK{
V.~Fateev, A.~B.~Zamolodchikov and A.~B.~Zamolodchikov,
``Boundary Liouville field theory. I: Boundary state and boundary  two-point
function,''
arXiv:hep-th/0001012.
}

\lref\TeschnerMD{
J.~Teschner,
``Remarks on Liouville theory with boundary,''
arXiv:hep-th/0009138.
}

\lref\ZamolodchikovAH{
A.~B.~Zamolodchikov and A.~B.~Zamolodchikov,
``Liouville field theory on a pseudosphere,''
arXiv:hep-th/0101152.
}

\lref\RibaultSS{
S.~Ribault and V.~Schomerus,
``Branes in the 2-D black hole,''
JHEP {\bf 0402}, 019 (2004)
[arXiv:hep-th/0310024].
}

\lref\DixonCG{
  L.~J.~Dixon, M.~E.~Peskin and J.~D.~Lykken,
  ``N=2 Superconformal Symmetry And SO(2,1) Current Algebra,''
  Nucl.\ Phys.\ B {\bf 325}, 329 (1989).
}

\lref\AhnTT{
C.~Ahn, M.~Stanishkov and M.~Yamamoto,
``One-point functions of N = 2 super-Liouville theory with boundary,''
Nucl.\ Phys.\ B {\bf 683}, 177 (2004)
[arXiv:hep-th/0311169].
}

\lref\IsraelJT{
D.~Israel, A.~Pakman and J.~Troost,
``D-branes in N = 2 Liouville theory and its mirror,''
arXiv:hep-th/0405259.
}

\lref\AhnQB{
C.~Ahn, M.~Stanishkov and M.~Yamamoto,
``ZZ-branes of N = 2 super-Liouville theory,''
JHEP {\bf 0407}, 057 (2004)
[arXiv:hep-th/0405274].
}

\lref\hoso{
K.~Hosomichi,
``N = 2 Liouville theory with boundary,''
arXiv:hep-th/0408172.
}

\lref\IsraelFN{
D.~Israel, A.~Pakman and J.~Troost,
``D-branes in little string theory,''
arXiv:hep-th/0502073.
}

\lref\BigazziMD{
  F.~Bigazzi, R.~Casero, A.~L.~Cotrone, E.~Kiritsis and A.~Paredes,
  ``Non-critical holography and four-dimensional CFT's with fundamentals,''
  JHEP {\bf 0510}, 012 (2005)
  [arXiv:hep-th/0505140].
}

\lref\CaseroPT{
  R.~Casero, C.~Nunez and A.~Paredes,
  ``Towards the string dual of N = 1 SQCD-like theories,''
  Phys.\ Rev.\ D {\bf 73}, 086005 (2006)
  [arXiv:hep-th/0602027].
}

\lref\BilalUH{
A.~Bilal and J.~L.~Gervais,
``New Critical Dimensions For String Theories,''
Nucl.\ Phys.\ B {\bf 284}, 397 (1987).
}

\lref\BilalIA{
A.~Bilal and J.~L.~Gervais,
``Modular Invariance For Closed Strings At The New Critical Dimensions,''
Phys.\ Lett.\ B {\bf 187}, 39 (1987).
}

\lref\WittenSC{
E.~Witten,
``Solutions of four-dimensional field theories via M-theory,''
Nucl.\ Phys.\ B {\bf 500}, 3 (1997)
[arXiv:hep-th/9703166].
}

\lref\BarsSR{
I.~Bars and K.~Sfetsos,
``Conformally exact metric and dilaton in string theory on curved
space-time,''
Phys.\ Rev.\ D {\bf 46}, 4510 (1992)
[arXiv:hep-th/9206006].
}

\lref\TseytlinMY{
A.~A.~Tseytlin,
``Conformal sigma models corresponding to gauged Wess-Zumino-Witten
theories,''
Nucl.\ Phys.\ B {\bf 411}, 509 (1994)
[arXiv:hep-th/9302083].
}

\lref\KutasovFG{
  D.~Kutasov, K.~Okuyama, J.~w.~Park, N.~Seiberg and D.~Shih,
  ``Annulus amplitudes and ZZ branes in minimal string theory,''
  JHEP {\bf 0408}, 026 (2004)
  [arXiv:hep-th/0406030].
}

\lref\BertoliniGG{
  M.~Bertolini, P.~Di Vecchia, G.~Ferretti and R.~Marotta,
  ``Fractional branes and N = 1 gauge theories,''
  Nucl.\ Phys.\ B {\bf 630}, 222 (2002)
  [arXiv:hep-th/0112187].
}

\lref\Teschner{
  J.~Teschner,
  ``On boundary perturbations in Liouville theory and brane dynamics in
  noncritical string theories,''
  JHEP {\bf 0404}, 023 (2004)
  [arXiv:hep-th/0308140].
}

\lref\GiveonSR{
  A.~Giveon and D.~Kutasov,
  ``Brane dynamics and gauge theory,''
  Rev.\ Mod.\ Phys.\  {\bf 71}, 983 (1999)
  [arXiv:hep-th/9802067].
}

\lref\bianchi{
M.~Bianchi, G.~Pradisi and A.~Sagnotti, Nucl.\ Phys.\ B {\bf B376}, (1992)  365}

\lref\calnap{
 C.~G.~.~Callan, C.~Lovelace, C.~R.~Nappi and S.~A.~Yost,
 ``Adding Holes And Crosscaps To The Superstring,''
 Nucl.\ Phys.\ B {\bf 293}, 83 (1987).
}

\lref\FNP{A.~Fotopoulos, V.~Niarchos and N.~Prezas,
 ``D-branes and SQCD in Non-Critical Superstring Theory,''
 [arXiv:hep-th/0504010].}

\lref\diVec{
P.~Di Vecchia, M.~Frau, I.~Pesando, S.~Sciuto, A.~Lerda and R.~Russo,
``Classical p-branes from boundary state,''
Nucl.\ Phys.\ B {\bf 507}, 259 (1997)
[arXiv:hep-th/9707068].
}

\lref\eguchi{
T.~Eguchi and Y.~Sugawara,
``Modular bootstrap for boundary N = 2 Liouville theory,''
JHEP {\bf 0401}, 025 (2004)
[arXiv:hep-th/0311141].
}

\lref\gaberdiel{
M.~R.~Gaberdiel,
``Lectures on non-BPS Dirichlet branes,''
Class.\ Quant.\ Grav.\  {\bf 17}, 3483 (2000)
[arXiv:hep-th/0005029].
}

\lref\janbranes{
D.~Israel, A.~Pakman and J.~Troost,
``D-branes in N = 2 Liouville theory and its mirror,''
arXiv:hep-th/0405259.
}

\lref\kutsei{
D.~Kutasov and N.~Seiberg,
``Noncritical Superstrings,''
Phys.\ Lett.\ B {\bf 251} (1990) 67.
}

\lref\li{
 M.~Li,
 ``Boundary States of D-Branes and Dy-Strings,''
 Nucl.\ Phys.\ B {\bf 460}, 351 (1996)
 [arXiv:hep-th/9510161].
}

\lref\sameer{
S.~Murthy,
``Notes on non-critical superstrings in various dimensions,''
JHEP {\bf 0311}, 056 (2003)
[arXiv:hep-th/0305197].
}

\lref\ouyang{
 I.~R.~Klebanov, P.~Ouyang and E.~Witten,
 ``A gravity dual of the chiral anomaly,''
Phys.\ Rev.\ D {\bf 65}, 105007 (2002) [arXiv:hep-th/0202056].
}

\lref\polchinski{J.~Polchinski, ``String theory, Vol. I, II'', Cambridge University Press (1998).}

\lref\ribschom{
  S.~Ribault and V.~Schomerus,
  ``Branes in the 2-D black hole,''
  JHEP {\bf 0402}, 019 (2004)
  [arXiv:hep-th/0310024].
}

\lref\yost{
 S.~A.~Yost,
 ``Bosonized Superstring Boundary States And Partition Functions,''
 Nucl.\ Phys.\ B {\bf 321}, 629 (1989).
}

\lref\farkas{H.~M.~Farkas and I.~Kra, ``Theta constants, Riemann surfaces and the modular group'', Graduate studies in mathematics, Vol.37. Amer.\ Math.\ Soc.\ }

\lref\tatatheta{D.~Mumford, ``Tata Lectures on Theta''}

\lref\mizo{S.~Mizoguchi, ``Modular invariant critical superstrings on four-dimensional Minkowski space $\times$ two-dimensional black hole'', JHEP {\bf 0004} 14 (2000) [arXiv:hep-th/0003053].}

\lref\bilal{
A.~Bilal and J.~L.~Gervais, ``New critical dimensions for string theories'', Nucl.\ Phys.\ B {\bf 284}, 397 (1987). }

\lref\absteg{
M.~Abramowitz and I.~Stegun, ``Handbook of Mathematical Functions''.
}

\lref\gepner{
 D.~Gepner, ``Lectures On N=2 String Theory,''
PUPT-1121 {\it Lectures at Spring School on Superstrings, Trieste, Italy, Apr 3-14, 1989} }

\lref\Ashok{
  S.~K.~Ashok, S.~Murthy and J.~Troost,
  ``D-branes in non-critical superstrings and minimal super Yang-Mills in
  various dimensions'',  [arXiv:hep-th/0504079].
}

\lref\KlebanovJJ{
I.~R.~Klebanov, J.~M.~Maldacena and C.~B.~Thorn,
``Dynamics of flux tubes in large N gauge theories,''
arXiv:hep-th/0602255.
}

\lref\PonsotGT{
  B.~Ponsot, V.~Schomerus and J.~Teschner,
  ``Branes in the Euclidean AdS(3),''
  JHEP {\bf 0202}, 016 (2002)
  [arXiv:hep-th/0112198].
}

\lref\RibaultPQ{
  S.~Ribault,
  ``Discrete D-branes in AdS(3) and in the 2d black hole,''
  arXiv:hep-th/0512238.
}

\lref\MartinecKA{
  E.~J.~Martinec,
  ``The annular report on non-critical string theory,''
  arXiv:hep-th/0305148.
}

\lref\AshokXC{
  S.~K.~Ashok, S.~Murthy and J.~Troost,
  ``Topological cigar and the c = 1 string: Open and closed,''
  JHEP {\bf 0602}, 013 (2006)
  [arXiv:hep-th/0511239].
}

\lref\SenNF{
  A.~Sen,
  ``Tachyon dynamics in open string theory,''
  Int.\ J.\ Mod.\ Phys.\ A {\bf 20}, 5513 (2005)
  [arXiv:hep-th/0410103].
}

\lref\MukhiZB{
  S.~Mukhi and C.~Vafa,
  ``Two-dimensional black hole as a topological coset model of c = 1 string
  theory,''
  Nucl.\ Phys.\ B {\bf 407}, 667 (1993)
  [arXiv:hep-th/9301083].
}

\lref\NakamuraSM{
  S.~Nakamura and V.~Niarchos,
  ``Notes on the S-matrix of bosonic and topological non-critical strings,''
  JHEP {\bf 0510}, 025 (2005)
  [arXiv:hep-th/0507252].
}

\lref\LeighEP{
  R.~G.~Leigh and M.~J.~Strassler,
  ``Exactly marginal operators and duality in four-dimensional N=1
  supersymmetric gauge theory,''
  Nucl.\ Phys.\ B {\bf 447}, 95 (1995)
  [arXiv:hep-th/9503121].
}

\lref\ElitzurHC{
  S.~Elitzur, A.~Giveon, D.~Kutasov, E.~Rabinovici and A.~Schwimmer,
  ``Brane dynamics and N = 1 supersymmetric gauge theory,''
  Nucl.\ Phys.\ B {\bf 505}, 202 (1997)
  [arXiv:hep-th/9704104].
}

\lref\ShifmanUA{
  M.~A.~Shifman,
  ``Nonperturbative dynamics in supersymmetric gauge theories,''
  Prog.\ Part.\ Nucl.\ Phys.\  {\bf 39}, 1 (1997)
  [arXiv:hep-th/9704114].
}

\lref\DouglasBN{
  M.~R.~Douglas,
  ``Branes within branes,''
  arXiv:hep-th/9512077.
}

\lref\SeibergPQ{
  N.~Seiberg,
  ``Electric - magnetic duality in supersymmetric nonAbelian gauge theories,''
  Nucl.\ Phys.\ B {\bf 435}, 129 (1995)
  [arXiv:hep-th/9411149].
}
\lref\BeasleyZP{
C.~E.~Beasley and M.~R.~Plesser,
JHEP {\bf 0112}, 001 (2001)
[arXiv:hep-th/0109053].
}
\lref\FengBN{
B.~Feng, A.~Hanany, Y.~H.~He and A.~M.~Uranga,
JHEP {\bf 0112}, 035 (2001)
[arXiv:hep-th/0109063].
}
\lref\CachazoSG{
F.~Cachazo, B.~Fiol, K.~A.~Intriligator, S.~Katz and C.~Vafa,
Nucl.\ Phys.\ B {\bf 628}, 3 (2002)
[arXiv:hep-th/0110028].
}

\lref\BerensteinFI{
  D.~Berenstein and M.~R.~Douglas,
  ``Seiberg duality for quiver gauge theories,''
  arXiv:hep-th/0207027.
}

\lref\GiveonPX{
A.~Giveon and D.~Kutasov,
``Little string theory in a double scaling limit,''
JHEP {\bf 9910}, 034 (1999)
[arXiv:hep-th/9909110]
\semi 
A.~Giveon, D.~Kutasov and O.~Pelc,
``Holography for non-critical superstrings,''
JHEP {\bf 9910}, 035 (1999)
[arXiv:hep-th/9907178].
}

\lref\HoriAB{
  K.~Hori, H.~Ooguri and Y.~Oz,
  ``Strong coupling dynamics of four-dimensional N = 1 gauge theories from  M
  theory fivebrane,''
  Adv.\ Theor.\ Math.\ Phys.\  {\bf 1}, 1 (1998)
  [arXiv:hep-th/9706082].
\semi
  E.~Witten,
  ``Branes and the dynamics of {QCD},''
  Nucl.\ Phys.\ B {\bf 507}, 658 (1997)
  [arXiv:hep-th/9706109].
%
\semi
  A.~Brandhuber, N.~Itzhaki, V.~Kaplunovsky, J.~Sonnenschein and S.~Yankielowicz,
  ``Comments on the M theory approach to N = 1 S{QCD} and brane dynamics,''
  Phys.\ Lett.\ B {\bf 410}, 27 (1997)
  [arXiv:hep-th/9706127].

}

\lref\AlishahihaYV{
  M.~Alishahiha, A.~Ghodsi and A.~E.~Mosaffa,
  ``On isolated conformal fixed points and noncritical string theory,''
  JHEP {\bf 0501}, 017 (2005)
  [arXiv:hep-th/0411087].
}

\lref\StrasslerQS{
  M.~J.~Strassler,
  ``The duality cascade,''
  arXiv:hep-th/0505153.
}

\lref\BarbonZU{
  J.~L.~F.~Barbon,
  ``Rotated branes and N = 1 duality,''
  Phys.\ Lett.\ B {\bf 402}, 59 (1997)
  [arXiv:hep-th/9703051].
}


\Title{\vbox{\baselineskip12pt\hbox{IC/2006/041}\hbox{LPTENS-06/23}
}}%
{\vbox{\centerline{D-branes in non-critical superstrings} 
\vskip8pt  
\centerline{and duality in $\CN=1$ gauge theories with flavor.}
}}

{\vskip -20pt\baselineskip 14pt   
\centerline{
Sameer Murthy$^{a,1}$ \footnote{}{$^1$\tt smurthy@ictp.it}  
and  
Jan Troost$^{b,2}$ \footnote{}{$^2$\tt troost@lpt.ens.fr} 
}  
  
\bigskip  

\centerline{\sl $^a$Abdus Salam International Center for Theoretical Physics}  
\centerline{\sl Strada Costiera 11, Trieste, 34014, Italy.}  
\smallskip  
\centerline{\sl $^b$Laboratoire de Physique Th\'eorique}  
\centerline{\sl Ecole Normale Sup\'erieure${*}$ \footnote{}{$*$Unit{\'e} mixte  du
CNRS et de l'Ecole Normale Sup{\'e}rieure, UMR 8549.}}
\centerline{\sl 24, Rue Lhomond, Paris 75005,  France.}
\bigskip\bigskip

\centerline{\bf Abstract}
We study  
D-branes in the superstring background $\IR^{3,1} \times SL(2,\IR)_{k=1}/U(1)$
which are extended in the cigar direction. Some of these branes are new. The
branes realize flavor in the four dimensional $\CN=1$ gauge theories on the D-branes 
 localized at the tip of the cigar. 
We study the analytic properties of the boundary conformal field theories on
these 
branes with respect to their defining parameter and find 
non-trivial monodromies in this parameter.
Through this approach, we gain a better understanding of the brane set-ups in ten
dimensions involving wrapped NS5-branes.
 As one application, using the boundary conformal field theory description of the electric and magnetic
D-branes, we can 
understand electric-magnetic (Seiberg) duality in $\CN=1$ SQCD 
microscopically in a string theoretic context. 
\noindent   
}

\Date{June 2006}

\newsec{Introduction and summary}

Non-critical superstring theory \kutsei\ is a part of the moduli space of
string theory compactifications with special properties. The dimension of
space-time can be lowered, the background contains fewer (super)symmetries, and there are 
fewer fields in the perturbative string spectrum. The
search for (linear dilaton) holography in lower dimensions, or the exploration of the full
moduli space of string theory compactifications  are sufficient motivation for
the study of these backgrounds. In this paper, we concentrate on their ability to
allow for the economical study of $\CN=1$ gauge theories with flavor (see
e.g. \refs{\KS , \KlebanovYA ,
\BigazziMD , \CaseroPT , \AlishahihaYV }  and \refs{ \FNP , \Ashok} ).

Many properties of  gauge theories have gained intuitive
interpretations via their embedding into string theory using D-branes. One such field where
 progress was made 
in understanding Seiberg duality in $\CN=1$ gauge theories 
 is in terms of brane set-ups  \refs{ \HananyIE , \ElitzurHC }. 
In particular the matching of the moduli spaces and the
chiral rings was achieved in terms of the embeddings of NS5-branes and
D-branes in flat space-times.

In this paper, we wish to develop the understanding of $\CN=1$ gauge theories
via their embedding into string theory further. 
In particular, the non-critical superstring theory set-up automatically takes
into account the backreaction of the NS5-branes (of mass $1/g_s^2$) on the closed string
background, in the (doubly scaled) near-horizon limit \GiveonPX . This allows us to
clearly separate effects due to the closed string NS5-brane solitons, and due
to the D-brane defects  (of mass $1/g_s$). Indeed, 
once the closed string backreaction due to the NS5-branes is taken care
off, we can study the D-branes in their presence using the boundary states
that code the boundary conformal field theory that the open strings give rise to.

We will argue in this paper that this provides a microscopic view on D-brane
set-ups and on Seiberg duality \SeibergPQ.  
In particular, we will see how the microscopy
 allows to clearly see the appearance and importance of the meson in the
magnetic dual, the phenomenon of Higgsing, and the  appearance of 
dangerously irrelevant operators in the field theory.
This involves a nice but detailed interplay between the boundary SCFT's on the worldsheet, the physics of the gauge theory and the closed string physics in these highly curved backgrounds. In the rest of this section, we summarize the flow of the main ideas in the paper.

We study the type IIB $d$ dimensional closed
non-critical superstring \kutsei\ background 
 $\IR^{d-1,1} \times
SL(2,R)/U(1)$. The factor $SL(2)_{k}/U(1)$ is a Kazama-Suzuki 
coset 
superconformal field theory \KazamaQP , 
which has a mirror description as the $\CN=2$ Liouville theory. The
supercoset has central charge $c=3+{6 \over k} = 3(1+Q^{2})$, where $Q$ is the
slope of the linear dilaton theory that the conformal field theory asymptotes to. We 
shall focus on the level $k=1$ which comes with a corresponding
four-dimensional ($d=4$) flat space
factor. 

We shall study D-branes which fill the flat space $\IR^{3,1}$. The low energy theories on
 the worldvolume of these D-branes are the four-dimensional minimally
 supersymmetric gauge theories.\foot{The branes which are Dirichlet in some of
 the flat directions are also interesting, they give rise to non-perturbative
 effects like domain
 walls and instantons \Ashok\ in the gauge theories.} The field content and
 interactions of the 
gauge theory depend on the profile of the brane in the cigar. 
The branes on the cigar are of two types, localized and extended\foot{These are similar to the ZZ \ZamolodchikovAH\ and FZZT  \refs{\FateevIK , \TeschnerMD}
branes in Liouville theory; in fact the relation between them is more than an analogy \AshokXC .}. The
localized branes are based on the identity representation and live at the tip of the cigar. They realize pure super Yang-Mills as their
world-volume theories. These were analyzed in \FNP\ and \Ashok\ independently (see
 also \eguchi). 
Here, we shall study further the
extended branes which realize flavor in $\CN=1$ gauge theories \refs{ \FNP , \KlebanovJJ }. 

The extended branes obey to a good approximation Neumann boundary conditions in the  weak-coupling region. As we go towards the strong coupling region, a worldvolume potential develops and they dissolve away. These branes are defined semi-classically by a complex parameter $\mu_{B}$ whose absolute value indicates how far they extend into the region near the tip. Quantum mechanically, they are better described by the parameters $(J,M)$ which are the labels of the $SL(2)$ representations from which they descend. 

For the continuous representation $(J=\half - iP, M=\half)$, the branes
 introduce quarks and anti-quarks into the gauge theory  \FNP . The mass
 of the quarks is set by the parameter $P^{2}$, and vanishes at $P^{2}=0$. For
 other values of $(J,M)$, the corresponding boundary states set down in
\refs{\eguchi ,\janbranes , \AhnTT } were studied in detail in \hoso . 
One aspect which was not very clear about these branes was whether they have a
 well-defined unitary  self-overlap.\foot{It was noticed \eguchi\ that for
 real values of $J \in ({1 \over 4}, \half$], the self-overlaps are
 well-defined.}. We find here that one can indeed understand  these branes
 systematically following the ideas of \TeschnerMD\ and there do exist branes
 with a unitary spectrum for values of $J$ in the range $0 \le J \le \half$. 

These branes have a semiclassical interpretation of turning on a worldvolume
 two form flux in the region near the tip of the cigar. At $J=0$, this value
 of the flux reaches a critical value and localizes at the tip, corresponding
 to forming a localized brane. In the quantum theory, this is seen in a
 relation between the boundary states which expresses the localized identity boundary state
 as a difference between the
 $J=M=0$ and the $J=M=\half$ boundary state. This phenomenon is similar to that in
 Liouville theory \MartinecKA ; 
 the  $\CN=2$ Liouville theory interprets this relation 
 as the localized brane being ``a brane inside a brane'' \DouglasBN\ 
{\it i.e.} built of gauge fields living on the extended
 brane. 

The quantum description of the branes 
 as boundary states  contains much more information. 
 The $J=M=0$ brane introduces flavors in the fundamental and anti-fundamental representation, 
 along with a gauge field into 
 the gauge theory on the localized branes. 
 Its self overlap contains, among other modes a localized mode with the quantum numbers
 of a meson! 
 
  We thus develop a picture of the Higgs mechanism in this theory
 as the following: the $J=\half$ flavor brane combines with the color brane to
 give the $J=0$ flavor brane. 
 The $J=0$ flavor brane thought of as a single object preserves only one combination of the two $U(1)$'s rotating the two branes independently. It 
 sits at the origin of the Higgs branch of the theory and realizes in its 
 spectrum the meson as a  Nambu-Goldstone boson of the broken symmetry.

From what we said above about the relation between the three branes (all of
 positive tension), it might seem like the $J=\half$ brane and the identity
 brane should be thought of (outside the Higgs branch) as the elementary
 branes. However, this turns out to be so because of a particular
 choice
 in orientation of the branes with respect to the closed string background\foot{The relative
 orientation arises while specifying the sine-Liouville potential
 to define the worldsheet theories and is an well-defined notion 
 because of the lack of further transverse directions, as is familiar from
 set-ups with
 anomalous creation of branes. }.
This change of orientation is implemented in the theory 
by changing the phase of the bulk parameter $\mu$ which multiplies the $\CN=2$ Liouville potential. 

The operation $\mu \to -\mu$ affects all the states in the theory including
the branes, and the net effect is that the new color brane and the $J=0$ brane
become elementary and the $J=\half$ brane can be expressed as a sum of these
two. Following this phenomenon onto the low energy theory realized on the
branes leads us to the electric-magnetic duality of Seiberg \SeibergPQ .  This
type of rearrangement of the basis of boundary states leading to Seiberg
duality was observed
at the level of the charge lattice 
in \refs{\BeasleyZP , \FengBN , \CachazoSG ,  \BerensteinFI } for 
quiver gauge theories. 
With an explicit worldsheet 
description of our theories and their branes at hand, we can 
lift these statements to the exact 
boundary states which includes the coupling to 
all the closed string excitations of the theory.

The boundary state description has one more piece of easily computable, useful
 information. It tells us how the closed string background backreacts to the
 presence of the source. It has been argued [see e.g. \BertoliniGG ] that the
 profile of the backreaction encodes the running of the coupling constant in
 the field theory. Such considerations in this case tell us that the gauge
 theory in question has indeed the spectrum of $\CN=1$ SQCD, but also suggest
 that there is an interaction  quartic in the quark superfields.\foot{This is
 also consistent with the lack of a chiral flavor symmetry in the theory.}
 This theory can be thought of as a softly broken $\CN=2$ theory which is
 sensible
 from the viewpoint of 5-brane embeddings in string theory as we shall explain.

The layout of the rest of the paper is as follows: In section \thebranes , we
 summarize the construction of the various boundary states on the cigar, and
 then present the addition formulas. In section \semiclass , we develop a
 semiclassical understanding of the various branes and their addition
 relations, and then present the transformation properties under the change in
 phase of $\mu$ referred to above. In section \overlap , we present the
 self-overlaps of the various branes, paying attention to the new ones -- we
 show the appearance of the localized modes, and explain their role in the
 gauge theory.  We then address the question of mutual supersymmetry and
 present a list of branes which preserve the same set of supersymmetries. In
 section \tend , we explain the relation between our theories and the more
 familiar brane set-ups in ten dimensions and point out where the exact
 treatment is not just
 pleasing but necessary. 

Having thus assembled all the ingredients and intuitions, we put them all
 together  in section \duality\ and read off as a consequence the duality in
 the low energy theory. Then, in section \bdryRG , we begin to develop an
 understanding of the processes as an RG flow in the boundary CFT. In section
 \quartic , we probe in more detail the interactions in the gauge theory under
 consideration. We compute a first order backreaction of the branes onto the
 closed string background and show that the anomalies of a classical $U(1)$
 global symmetry of the theory is encoded in the backreaction onto the RR
 axion. From the backreaction onto the NSNS sector, we argue that the gauge
 theory under consideration is really $\CN=1$ SQCD with a quartic coupling of
 the quarks. In section \final , we make some final remarks and point out
 open issues.

\newsec{The branes on the cigar}
\seclab\thebranes 

The boundary states on the $\CN=2$ supercoset $SL(2,R)/U(1)$ 
can be  classified as $A$ type or $B$ type according to the
gluing of the left-moving $\CN=2$ algebra to its right-moving counterpart. 
When tensored with space filling
boundary states in the flat space $\IR^{3,1}$,  the B-branes in the cigar 
are supersymmetric in the type IIB closed superstring
theory. Since the $U(1)$ $R$-current of the $\CN=2$ superconformal algebra on
the worldsheet contains a term that translates 
the chiral boson corresponding to the angular direction $\theta$ of the cigar,
the B-type boundary condition always leads to Neumann boundary conditions in
the angular direction $\theta$, {\it i.e.} the B-branes conserve the momentum around the cigar. If a
brane extends to the asymptotic weak coupling region where a semi-classical
notion is valid, we can say that the supersymmetric branes
 wrap the circle of the cigar. 
We will sometimes use the picture of the T-dual type IIA superstring theory on the $\CN=2$
Liouville theory with a momentum sine-Liouville condensate in which the
supersymmetric  A-branes conserve the 
winding around the cylinder.

A large class of consistent boundary states has
been presented in a clear manner in \hoso , summarizing and extending earlier
 work \refs{ \eguchi , \janbranes ,\AhnTT }. 
The consistency conditions on the boundary states include the requirement that the
spectrum of boundary operators be sensible (e.g. has a positive density of
states), as well as a factorization equation which leads to a so-called
shift-equation,
analyzed in \refs{\PonsotGT , \hoso}. A complete check of the consistency requirements
on these non-rational boundary
conformal field theories remains to be performed.

It is crucial \eguchi\ that the set of B-branes contains an identity brane that by
definition
has only  the identity
representation in its 
open string spectrum. Its overlap with other
B-branes contains (as a consequence of a generalized Verlinde formula \JegoTA 
) 
only the single representation
characterizing those branes. Thus, we can review first the relevant characters of
the representations of the $\CN=2$ superconformal algebra,  then recall the
associated D-branes.

The relevant representations 
are actually representations of the $\CN=2$ superconformal
 algebra extended by a spectral flow operation, leading to extended characters
 {\it i.e.} characters summed over spectral flow orbits \eguchi . It is convenient
to label these characters in terms of quantum numbers associated to a parent
supersymmetric
$SL_{2}$ theory. The characters are then
 labeled by the numbers $(J,M)$, parameterizing their parent $SL_{2}$
 Casimir and compact $U(1)$ charge. Equivalently they can be labeled by their $\CN=2$
 conformal dimension and R-charge $(h,Q)$, where these are given in
 terms of the quantum numbers $(J,M)$ by the formulas $h=-{J(J-1) \over k}+{M^{2} \over k}$
and $Q={2M \over k}$ in the NS sector. 
Another convenient parameterization, natural for continuous representations of 
$SL_{2}$, is given by the formula $J=\half - iP$, where $P$ is a momentum variable.
For a detailed discussion of these issues see \refs{\DixonCG , \janbranes }.

In order not to clutter the formulas, in the following we immediately restrict
to our case of interest, namely the level $k$ of the coset equals
one.
For $k=1$, the relevant extended characters in the NS sector are 
labeled by a charge $Q \in \{0,\pm 1\}$. We distinguish representations according to
whether they descend from continuous or discrete representations,
or the identity representation
in the parent $SL_{2}$ theory: 
\eqn\Exchar{\eqalign{
& \qquad {{\rm \bf Continuous:}}  \quad J=\half - iP, P \in \IR^{+}, M \in
\{0,\half \} \cr
& Ch_{cont}^{NS}(h,Q;\tau,z)  = q^{h- {Q^{2} +1 \over 4}} \sum_{m \in \IZ} q^{(m+{Q \over 2})^{2}} y^{2m+Q}   {\vartheta_{00} (\tau,z)\over \eta^{3}(\tau)} \cr
&  \qquad {\rm \bf Discrete:}  \quad  J=|M|  \cr
& Ch_{disc}^{NS}(Q;\tau,z)  = q^{-{1\over 4}} \sum_{m \in \IZ}{q^{m^{2}+|Q|m+{|Q| \over 2}} y^{sgn(Q)(2m+|Q|)}    \over 
(1+y^{sgn(Q)} q^{m+\half})} {\vartheta_{00} (\tau,z)\over \eta^{3}(\tau)} \cr
& \qquad {{\rm \bf Identity}}:  \quad J=M=0 \cr
& Ch_{Id}^{NS}(\tau,z)  = q^{-{1\over 4}} \sum_{m \in \IZ}{(1-q)q^{m^{2}+m-\half} y^{2m+1}    \over 
(1+y q^{m+\half})(1+yq^{m-\half})} {\vartheta_{00} (\tau,z)\over \eta^{3}(\tau)}. \cr
}}
Note that the continuous characters depend on $|Q|$ only. The characters
form a representation of the modular group \eguchi\ .
The characters in the other sectors $\T NS, R, \T R$ can be found
by $\CN=2$ spectral flow (see e.g.\refs{ \eguchi , \janbranes}).

Corresponding to each of these characters, there is a 
consistent boundary state \hoso. These boundary states are constructed in the
following manner.
 We first define the Ishibashi states:
\eqn\ishi{
\bbra{p,Q} e^{- \pi T H^{cl}}e^{i \pi z(J+\T J)} \kket{p',Q'} = 2 \pi \left( \delta(p-p') + R(p) \delta(p+p') \right)\delta_{\IZ_{2}} (Q,Q') Ch^{NS} (p, Q, i T, z).
}
Then the identity brane corresponds to the one-point functions\foot{We have included here the dependence on the bulk interaction coefficient
$\nu$ which we assume to be real for now. We
 discuss the physics of its phase later.}:
\eqn\idbrane{\eqalign{
\ket{B;Id} & = \int_{-\infty}^{\infty} {dp' \over 2 \pi} \sum_{Q' \in \IZ_{2}} \Psi_{Id} (p',Q') \kket{p',Q'} \cr
& \Psi_{Id}(p',Q') =  {     ({\pi \over 2})^{1 \over 2}} 
\nu^{i p'}
{\G(\half + {Q' \over 2} + {ip'})
\G(\half - {Q' \over 2} + {ip' }) \over \G(i 2 p') \G(1+i 2 p') }. \cr
}} 
It can be checked that it has only the identity character in its self-overlap.

\ndt The brane associated to the continuous representation has a Cardy state:
\eqn\contbrane{\eqalign{
\ket{B; cont, P,Q} & = \int_{-\infty}^{\infty} {dp' \over 2\pi} \sum_{Q' \in \IZ_{2}} \Psi_{cont} (p',Q') \kket{p',Q'} \cr
\Psi_{cont}(p',Q') &
 = (2 \pi)^{1 \over 2}  \nu^{i p'}   \cos{(4 \pi P p')}
{\G(1-i 2 p') \G(-i 2 p')    \over \G({1 \over 2} -i p'  +{Q' \over 2}) \G({1
\over 2}-i  p'- {Q' \over 2}) }
e^{\pi i Q Q'}. \cr
}}
The brane carries only the continuous representation in its overlap with the
identity brane. 
Finally, the brane carrying only the discrete representation in its overlap with the 
identity brane can be described by the formulas:
\eqn\discbrane{\eqalign{
\ket{B; disc, Q} & = \int_{-\infty}^{\infty} {dp' \over 2 \pi} \sum_{Q' \in \IZ_{2}} \Psi_{disc} (p',Q') \kket{p',Q'} \cr
\Psi_{disc}(p',Q') & = 
i (8 \pi)^{- \half } \nu^{i p'} e^{i \pi Q Q'} \G (1-2 i p') \G(-2ip') \times 
\cr
& 
\left( e^{ i \pi ({Q' \over 2} -\half +i p')} e^{-i 4 \pi p' p} {\G(\half+ip'+{Q' \over 2} ) \over
\G(\half-ip'+{Q' \over 2} ) }
- e^{ i \pi ({Q' \over 2}  +\half -i p')} e^{i 4 \pi p' p} {\G(\half+ip'-{Q' \over 2} ) \over
\G(\half-ip'-{Q' \over 2}) }
\right) \cr
}}
This is the anti-chiral brane in \hoso. There is another discrete brane
corresponding to the chiral brane in \hoso. 
The one-point functions in the other sectors $ \T NS, R, \T R$ can be obtained
by spectral flow.
It was checked in \hoso, and further in \RibaultPQ\  that the branes listed
above obey a bulk-boundary factorization constraint (i.e. the shift equation).  

There are restrictions on the values
 of the parameters $(J,M)$ labeling the Cardy states. 
 The open string
 spectrum\foot{In this particular set-up 
a light-cone gauge choice in some
extra flat directions is possible.} 
allows only for unitary representations to appear in the cigar brane overlap.
Moreover, we need to demand mutual consistency of the boundary states as well
 as with the bulk
 spectrum of the theory. 
 Our strategy  will be to 
 consider the branes to be analytic functions of the
 parameters $(J,M)$ in some of the calculations that follow,
and take care to impose all restrictions that follow 
from consistency  and unitarity.

\subsec{Addition relations obeyed by branes}
For the level $k=1$, the extended characters thought of as 
analytic functions of $(J,M)$ obey two identities \eguchi: 
\eqn\charaddn{\eqalign{
{\rm \bf Character \;} &{\rm \bf addition \; formulas}   \cr
    Ch_{cont}(h=\half,|Q|=1;\tau,z)  = &\; Ch_{disc}(Q=1;\tau,z) + Ch_{disc}(Q=-1;\tau,z)   \cr
    Ch_{cont}(h=0,Q=0;\tau,z)  = & \; Ch_{cont}(h=\half,|Q|=1;\tau,z)  +   Ch_{Id} (\tau,z). \cr
}}

\ndt What is even more powerful 
is that the corresponding branes obey related identities which
 can be checked by using the explicit one-point functions in formulas 
\idbrane, \contbrane, and \discbrane . Using these we obtain the important addition
relations between the branes: 
\eqn\braneaddn{\eqalign{
  {\rm \bf Brane \;} & {\rm \bf addition \; formulas}   \cr
    \ket{B;cont,h=\half,|Q|=1} & = \ket{B; disc, Q=1} +   \ket{B; disc, Q=-1}    \cr
 \ket{B;cont,h=|Q|=0} & =  \ket{B;cont,h=\half,|Q|=1} +  \ket{B;Id}. \cr
}}
In the next section, we shall  develop a semiclassical spacetime
understanding ($g_{s} \to 0$) of the various branes and in particular the
above two equations in  \braneaddn \foot{By the modular bootstrap reasoning, the brane
addition formula \braneaddn\ implies the character addition formula \charaddn
.}.  
We remark already that the second addition relation is
 similar to an equation for boundary states in bosonic $c=1$ theory\foot{Other
 precise relations between the branes of the two theories were written in \AshokXC .} \refs{\FZZ , \MartinecKA , \Teschner }. We
 will understand
 its implications to gauge theory physics in the following sections. 
In the rest of this section, we shall summarize the open 
string sigma-model ($\alpha' \to 0$) understanding  of the branes.

\subsec{The branes as defined by the Boundary cosmological constant}
\subseclab\bdrycc
Let us recall that the bulk theory is defined by a complex parameter, the
coefficient $\mu, \bar{\mu}$ of the $\CN=2$ Liouville coupling $\CL^{SL} = \mu \, \psi \t \psi \, e^{-{1 \over Q} (\rho +\t \rho + i (\theta - \t \theta ))} + c.c$ where we have used the asymptotic variables on the cigar. There is also the 
related parameter $\T \mu$,
the coefficient of the cigar interaction in terms of which the correlators can
be defined. For $\mu = \bar{\mu}$, the equation relating the coupling 
constants
is \hoso :
\eqn\mubulk{
(g_{s}^{tip})^{-2} = \mu^{2/k} = \T \mu {\G({1 \over k}) \over \G(1-{1 \over k})} \equiv \nu . 
}
The case of level $k=1$ needs a renormalization \Ashok\  similar to the Liouville
 theory at $b=1$ \McGreevyEP\ , as can be seen from the bulk tachyon
reflection amplitude. 
We have, with $k=1 + \epsilon$, 
\eqn\mubulkren{
(g_{s,ren}^{tip})^{-2}= \mu_{ren}^{2} =  \nu_{ren}; \quad \mu_{ren} = \mu \epsilon
}

\ndt On the boundary, the coupling constants are the cosmological constants
$\mu_{B}, \bar\mu_{B}$ and the non-chiral coupling $\T \mu_{B}$, 
 related to the brane labels $(J,M)$ by the equations \hoso :
\eqn\mubdry{\eqalign{
\mu_B &= ({2 k \mu \over \pi})^{1/2} \sin{(\pi (J-M))}
\cr
\bar{\mu}_B &= ({2 k \mu \over \pi})^{1/2} \sin{(\pi (J+M))}
\cr
\nu_{B} \equiv  \T \mu_{B} {\G({1 \over k}) \over \G(1-{1 \over k})} &=
 - {\mu \G({1 \over k}) \over 2 \pi } \cos{\left({\pi \over k} (2J+1)\right)} \cr
}}
For the level $k=1$ then, the renormalized cosmological constants (which enter the open
string amplitudes) are defined
 by the formulas:
\eqn\mubdryren{\eqalign{
\mu_{B} \epsilon^{1/2} \equiv \mu_{B,ren} &= ({2  \mu_{ren} \over \pi})^{1/2}
\sin{(\pi (J-M))}
\cr
\bar{\mu}_{B} \epsilon^{1/2} \equiv \mu_{B,ren} &= ({2  \mu_{ren} \over \pi})^{1/2}
\sin{(\pi (J+M))} \cr
\nu_{B} \epsilon \equiv \nu_{B,ren} & = - {\mu_{ren} \over 2 \pi }
\cos{\left(\pi (2J+1)\right)} = {\mu_{ren} \over 2 \pi } 
\cos{\left(2 \pi J \right)}. \cr
}}

\newsec{Semi-classics and covariance}
\seclab\semiclass
In this section, we discuss some of the semi-classical properties of the
D-branes
in the cigar, and the covariance properties of the boundary states under a
$Z_2$ operation.
\subsec{Semi-classical description of the branes}
We observe that the  D1-branes of sine-Liouville theory with momentum condensate (see
e.g. \FNP ) reach asymptotic infinity from two
angular directions, with an angular parameter
 $\theta_0$
 that
takes values in a full circle of length $ 2 \pi$. 
When we concentrate on the
 brane that stays fixed in the angular direction as it goes to the strongly
coupled region ($r=0$) however, we note that the NSNS
 sector one-point function is invariant under the operation
$\theta_0 \rightarrow \theta_0+ \pi$. This is directly associated to the fact
that the D1-branes have two legs, and that it allows for half-integer winding
open strings. 
In the full supersymmetric theory, we would moreover pick up a sign
in the RR term in the boundary state under this operation, due to the
 difference in orientation
of the resulting rotated D1-brane. In contrast, the localized D1-branes of
 \RibaultPQ\  (which we may 
think of as difference of (anti-)chiral
 branes in the nomenclature of \hoso ) are only invariant under the full $2 \pi$ rotation,
 showing that they have, in this sense, only a single leg (which in the case of
 these D1-branes is localized near the more strongly coupled region).

The picture we sketched above is T-dual to the
D2-branes on the cigar whose physics we can now more easily picture. For D2-branes, 
the differences discussed above
are reflected in having either a single-sheeted D2-brane, or a double-sheeted
D2-brane \FNP . The associated Wilson line on the D2-brane is either $2 \pi$ or
$\pi$ valued. Our discrete brane \discbrane\ which can be identified with 
 the D2-brane of \ribschom\  and  the anti-chiral
brane of \hoso\  is single sheeted. In contrast
the continuous brane consist of two oppositely oriented sheets (represented by
the sum of a chiral and an anti-chiral brane). 
Thus, we have given a geometrical picture for the first addition relation in \charaddn .

The picture we developed also explains why the single-sheeted discrete D2-branes
give rise in the full theory to a RR-tadpole that cannot be absorbed in the
background. It leads to an inconsistent bulk theory (for space-filling branes
in the flat directions). By open-closed string duality, 
should we add such a brane to the theory, we
would discover anomalous chiral matter in the massless open string sector. 
In contrast, the two-sheeted continuous brane cleverly cancels the potential RR-tadpole by
having two sheets of opposite orientation (nevertheless preserving
supersymmetry). See also \FNP .

The second addition relation shows that the difference between the two
particular two-sheeted space-filling branes under study is precisely the brane
localized at the tip. 
The brane $J=\half+iP$ extends from infinity to a certain distance determined by $P^{2}$ from the tip where it dissolves, as can be seen from the one-point functions \FNP . As $P =0$, it covers the whole cigar. On adding a localized brane at the tip, we get formally the $J=0$ brane. 

We can understand this ``addition'' by realizing that the localized brane
 sources a two-form flux on the extended brane. The $J=0$ brane admits a
 deformation where the cigar is still covered, but the distribution of the two
 form field changes. On turning on this deformation, the B-field spreads out
 from the tip and localizes in a ring at a small distance from the tip
 \IsraelFN . Semi-classically, we can identify this parameter to be $J \in
 [0,\half]$. The $J=0$ brane then is simply a point in this moduli space where
 the dissolved brane becomes point-like at the tip.

\subsec{Dependence  of the branes on the bulk interaction parameter $\mu$, and the action of $\mu \to -\mu$.}
In this subsection, we want to study how the closed string modes and the D-branes change under a 
change of phase of the bulk interaction parameter $\mu$, and in particular under  $\mu \to -\mu$.
The action $\mu \to -\mu$ is not a symmetry of the $\CN=2$ theory as can be
 seen from the closed string interaction written in section \bdrycc . 
 However, the covariance of the theory
 tells us how the objects in the theory defined by $\mu$ are related to the
 objects in the theory with $-\mu$. From the form of the worldsheet action, we also see
 that the operation $\mu \to -\mu$ is equivalent to the action $(-)^{w}:
 \theta \to \theta +  \pi$ and $\T \theta \to \T\theta -  \pi$ on the
 left-moving and right-moving angular coordinate in the cigar
theory.  In the T-dual sine-Liouville picture, this is a rotation of the cylinder by $\pi$.

In the perturbative sector of the theory, closed string modes in the sector with odd values of asymptotic 
winding pick up a sign, and those with even values of winding do not.
The tachyon state of winding one which has a condensate in the theory picks up a sign under the operation.

Now we ask what is the action on the boundary states.  Firstly, there is an implicit 
dependence through the coupling to the closed string modes. We can use the
same Ishibashi basis \ishi\ as before and keep track of the 
sign dependence as a phase in the one-point function.
The identity brane depends only on the closed string parameter $\mu$, and no
other
 intrinsic parameter, and so the same should be true about its corresponding
one point function. We can write 
the $\mu$ dependence as: 
\eqn\idbranemu{
\Psi_{Id}(p',Q';\mu) =  \mu^{ip'  - {Q' \over 2}} \bar{\mu}^{ip'  + {Q' \over 2}} ({\pi \over 2})^{\half} {\G(\half + {Q' \over 2} + {ip'}) \G(\half - {Q' \over 2} + {ip' }) \over \G(i 2 p') \G(1+i 2 p') }
}
which is consistent with the bulk reflection amplitude \eguchi . Using
the one-point functions
\contbrane , \discbrane , we can also write a similar equation for the $\mu$
dependence of the
 extended branes.

In addition, the action of the rotation   $(-1)^{w}$ could induce explicit changes of sign due to the full boundary state having a well-defined charge under the above mentioned symmetry $\mu \to -\mu$ accompanied by $(-1)^{w}$. To test this, we look at the coupling of the on-shell tachyon winding mode with the localized and extended branes. From \idbrane , \contbrane , we see that the former couples (with an infinite coefficient (see \Ashok )) and the extended brane  $\ket{B;J=M=\half}$ has vanishing coupling. We deduce the following transformations under $\mu \to -\mu$:
\eqn\mutominus{\eqalign{
& \ket{B;Id;\mu}_{NS} \to  - \ket{B;Id;-\mu}_{NS} \cr
& \ket{B;J=M=\half;\mu}_{NS} \to   \ket{B;J=M=\half; -\mu}_{NS} \cr
}}
Note that the brane $\ket{B;J=M=0}$ which is a linear combination of the above
two does 
not have a fixed transformation property under the operation.

It is important to note that the tension of the new localized brane 
$(- \ket{B;Id;-\mu}_{NS})$ is 
positive -- this can be seen by the computation of \Ashok\ where the tension was worked out to be proportional to $\nu_{bulk}^{\half} = \mu_{bulk}$. A change in the overall sign of the boundary state  and a change in sign of $\mu$ together give a factor of unity. At an intuitive level, we have that the  tension of the brane localized at the tip is proportional to ${g_{s}^{tip}}^{-1} = \mu_{bulk}$ \mubulkren . A change in its sign forces a change in sign of the positive tension boundary state.

Let us now discuss the RR sector of the boundary states. Under the rotation of the cylinder with a sine-Liouville potential by $\pi$,  the full state $\ket{B;J=M=\half}$ changes orientation with respect to the potential. This orientation does not make a difference for the NS sector of the boundary state, but it implies that the RR sector state changes sign with respect to the RR sector closed string fields on the cylinder (e.g. RR one-form in type IIA).\foot{One can instead compare the RR sectors of the above brane and the localized brane -- the localized brane $\ket{B;Id}$ rotates along with the potential and the relative orientation of the two branes  changes.} We will see later that this is consistent with the NS5-brane setup in ten dimensions. 
We write then the final equations describing the transformations of the branes under $\mu \to -\mu$:
\eqn\muBtominus{\eqalign{
& \ket{B;Id;\mu} \to  - \ket{B;Id;-\mu} \cr
& \ket{\bar{B;J=M=\half;\mu}} \to \ket{B;J=M=\half;-\mu}.\cr
}}

\newsec{The spectrum of open strings ending on various branes}
\seclab\overlap

In this section, we shall  address the issue of the
self-overlaps of the continuous branes, 
the overlaps between the localized and continuous branes,
and the mutual supersymmetries preserved by the various branes. 
At the end of the section, we will present
a summary of branes which are mutually supersymmetric and have a unitary 
spectrum in their overlaps. 

\subsec{The self overlap of the extended branes}
As mentioned earlier, we can formally define branes based on the continuous
and discrete characters with arbitrary values\foot{We shall impose the Seiberg
bound $J  \le \half $ following \Teschner .} of $(J,M)$. 
A necessary condition for consistency of the branes
is that  their self-overlap, and overlaps with other well-defined branes give
rise to unitary spectra.\foot{This problem was touched upon 
in \eguchi .} We shall focus on the continuous branes since we saw in the last
section that they do not have a Ramond-Ramond tadpole at 
 infinity. 
 
We choose a parameterization $J=\half - iP$, where $P$ is allowed to take
 complex values. The overlap between two branes can be found by expanding the
 two branes using the one-point
 functions \contbrane\ in the defining Ishibashi basis \ishi . One finds \eguchi,
after an exchange of order of integration to which we shall return shortly:
\eqn\contlap{\eqalign{
 e^{\pi {3 z^{2} \over T}} 
\bra{B;J_{1},M_{1}} e^{- \pi T H^{cl}}e^{i \pi z(J+\T J)} \ket{B;J_{2},M_{2}} 
= & \int_{-\infty}^{\infty}  {dp} \left[ \rho_{1}(p|J_{1},J_{2}) Ch(p,M_{2}-M_{1};it,z') \right. \cr
&\qquad \qquad + \left. \rho_{2}(p|J_{1},J_{2})  Ch(p,M_{2}-M_{1} +\half ;it,z') \right] \cr
}}
where the spectral densities $\rho_{i}$ are given by:
\eqn\specden{\eqalign{
& \rho_{1}(p | J_{1},J_{2})  =  \int_{0}^{\infty} {dp'} {\cos{(4 \pi p p')} \over \sinh^{2}{(2 \pi  p')} } \sum_{\epsilon_{i}=\pm 1} \cosh{\left(4\pi \left(\half + i \epsilon_{1} P_{1} + i \epsilon_{2} P_{2}\right) p' \right)}
\cr 
& \rho_{2}(p|J_{1},J_{2})  = 2 \int_{0}^{\infty} {dp'} {\cos{(4 \pi p p')} \over
 \sinh^{2}{(2 \pi  p')} } \sum_{\epsilon=\pm 1} \cosh{\left(4\pi \left( i
 P_{1} + i \epsilon P_{2}\right) p' \right)} .
\cr 
}}
For $P_{i} \in \IR^{+}$, these formulas are well-defined. For imaginary values of
 $P_{i}$ which we are interested in, corresponding to $0 \le J \le \half$, one has
 to be more careful. The $p'$ integral in \specden\  may generate additional
 divergences at $p' = \infty$, which can
 be eliminated by shifting the contour of $p$ integration in \contlap\  
{\it
 before exchanging the order of the integrals}  as explained in \TeschnerMD . 
Thereafter, one can freely exchange the integral and shift back the contour,
finding additional contributions to the brane spectrum.

 As a warmup, let us view how the above analysis
accords with the addition relation for the branes. From the addition relation, we
would expect a $J=M=0$ brane to have the same spectrum as a $J=M=\half$ brane, with
two extra continuous representations at $j=m=\half$ and another localized mode
corresponding to the identity character. This is in accord with the following
observation. The densities associated to the $J=0$ brane and the $J=\half$ brane
are related as follows:
\eqn\relationdensities{\eqalign{
\rho_{1}(p | J=0)  & = \rho_{1}(p | J=\half)
+ 4 \int_{0}^{+\infty} dp' \cos 4 \pi  p p' \cosh 2 \pi p'
\cr 
\rho_{2}(p | J=0)  & = \rho_{2}(p | J=\half)
+ 4 \int_{0}^{\infty} dp' \cos 4 \pi p p'.
\cr 
}}
These integrals are divergent if $(p,p')$ are real, but let us imagine having solved this 
problem by shifting the contour of $p$ in \contlap\ for the moment and proceed unhindered. 
We see that the $J=0$ self-overlap picks up a delta function contribution at $p=0$ from the second
density $\rho_2$, after $p$-integration, giving rise to one continuous character
at $j=\half$. The first density $\rho_{1}$, after integration over $p'$, contains two
poles, at $i p = \pm \half $. After shifting (back) the contour of integration of $p$ to the real axis, 
the integral on the real axis vanishes
due to anti-symmetry, and the pole at $ip= \half $  is picked up in the process,
contributing a continuous $j=0$ character. The latter splits into the second
$j=\half$ character and the identity character. Thus, we see that the annulus
spectrum is consistent with the brane addition relation.

We will now put all of this on a firm footing using the ideas of  \TeschnerMD.
We first shift the contour of the $p$ integration in the complex plane to make the 
integral \specden\ well-defined for imaginary $p_{i}$ as well. 
The spectral densities $\rho_{i}$ have divergent pieces independent of the
 boundary states coming from the region $p'=0$.
These can be subtracted off, and one can define relative spectral densities. 
We use the regularization of \FateevIK\ in terms of the q-Gamma function with $b=1$:
\eqn\qgamma{
\log S_{b}(x) = \int_{0}^{\infty} {dt \over t} \left[ {\sinh(Q-2x)t \over 2 \sinh(bt) \sinh(t/b)}  - {(Q/2-x) \over t} \right]; \quad Q=b+{1\over b}.
}
The convergent parts of the spectral densities are the following:
\eqn\specdenfin{\eqalign{
\rho_{1}(p | J_{1},J_{2})  & = {1 \over 8 \pi i} \sum_{\epsilon_{i}=\pm 1} \epsilon_{0} \p_{p} \ln \CS_{1}{\left(\half +i \epsilon_{0} p + i \epsilon_{1} P_{1} + i \epsilon_{2} P_{2} \right)}
\cr 
\rho_{2}(p | J_{1},J_{2})  & = {1 \over 4 \pi i} \sum_{\epsilon_{i}=\pm 1}  \p_{p} \ln \CS_{1}{\left(1 +i p + i \epsilon_{1} P_{1} + i \epsilon_{2} P_{2}\right) }.
\cr 
}}
Now, on shifting back the contour of integration of $p$, there is  a possibility of
 picking up additional contributions from the poles of the functions
 \specdenfin . For the self overlap, we have $J_{1}=J_{2}=:J$. 
Some details of this computation are given in appendix A, and we obtain the following
 results:

\vskip 0.2 cm

\ndt {\it Results of the self overlap computation for the extended branes:}
\item{1a.}  For $J>0$, the function $\rho_{2}(j)$ defined by \specden\ needs no shift of contour for its convergence. 
\item{1b.}
For the case $J=0$, a pole and a zero in the third and the fourth $S_{1}$ functions cancel each other and there is no extra pole. 
 There {\it is} however a delta function contribution to the function $\rho_{2}(j)$ at $j = \half - ip = \half $. This is due to a crossing of the  branch cut of the logarithm in \specdenfin . 
\item{2a.} For ${1 \over 4} < J \le \half$, the contour in \specden\ for the function $\rho_{1}(j)$ is well-defined, and there are no extra contributions to the spectrum.
\item{2b.} For $0<J \le {1 \over 4} $, there is a delta function contribution
to $\rho_{1}(j)$ at $j=\half$ for the same reason as above.
\item{2c.} For $J= 0 - \epsilon, \, \epsilon \ge 0$, 
there is a pole at $j= \half - ip = 0$ whose residue is unity.

\vskip 0.2 cm 

\ndt {\it Comments}:
\item{1.} The branes with $J\le 0$ have in their spectrum new extra localized  modes
contained in the continuous character $J=M=0$. Many of these branes ({\it e.g} $J<0$) have a
non-unitary spectrum in their overlap with the identity brane, as we shall see soon. 
\item{2.} The computation above was for the NS character. The presence of the
other three sectors will be dictated by supersymmetry.
\item{3.} The delta function in $\rho_{1}$ and $\rho_{2}$ also makes its appearance in the boundary Liouville theory, as is consistent with the addition relation in that context. In fact, the function $\rho_{2}$ is exactly the spectral density on the extended branes of Liouville theory with an appropriate change of variables.

\subsec{ Mutual Supersymmetry and the  GSO projection}
We have understood the various boundary states in the cigar SCFT from
different points of view.  Now we want to focus on the properties of the
branes in the full six-dimensional string theory. From this point on, we shall
discuss these branes, using the construction of \refs{ \Ashok , \eguchi } to complete
the cigar boundary states into the full D-brane boundary state. We need to make a GSO
projection in the open string channel consistent with the closed string spectrum. In
\Ashok , we discussed only one type of brane, using the boundary state
$\ket{B,Id}$. When there are no other branes present in the background, 
the two possible GSO projections are equivalent, giving rise to a brane
$\ket{D3}$ and an
anti-brane
$\ket{\bar{D3}}$. We shall denote the full extended branes and anti-branes 
 by $\ket{D5;J,M}$ and $\ket{\bar{D5; J,M}}$.

We are interested in the question of how much supersymmetry is preserved in
 the presence of one or more of the branes we have described. The closed string background 
 has $\CN=2$ Poincare supersymmetry in $d=4$, and has a $U(1)_{R}$ symmetry arising from the rotation of the cigar \sameer .  
  The localized brane $\ket{D3}$ and the anti brane $\ket{\bar{D3}}$ preserve half
 of the eight bulk supercharges. They do not preserve any of the same
 supercharges and a configuration of a $\ket{D3}$ and a $\ket{\bar{D3}}$ is
 non-supersymmetric and has a tachyon in
 the spectrum. 

The arguments for supersymmetry in  \Ashok\  (Appendix A) relied basically on
 the Neumann boundary condition for the $R$-current of the $\CN=2$ theory. All
 the branes in question here are $B$-branes and preserve the Neumann boundary
 condition, and so are half BPS by themselves. They all conserve the $U(1)_{R}$ symmetry.\foot{The boundary states are invariant under this symmetry, the backreaction onto the background causes this symmetry to be anomalous on the branes \Ashok .} 
 The question of mutual
 supersymmetry thus boils down to a GSO projection, which can be seen in the (non)vanishing of 
the annulus diagram between the various branes.

Let us first investigate the supersymmetry of the branes relative to the
$\ket{D3}$ brane. We note that the twisted NS and the R-sectors characters
follow by spectral flow from the NS sector.
After tensoring the flat space parts, we have the following two sets
of  branes  based on the continuous characters with vanishing 
overlap with the D$3$-brane (with $h=-J(J-1)+M^{2}$ arbitrary):
\eqn\susyqzero{\eqalign{
\bra{D3} e^{-TH_{cl}} \ket{D5; J,M=0}  & = \half {q^{h-{1 \over 4}} \over
\eta^{6}(\tau)}  \left[ \vartheta_{00} (2 \tau, 2z) \left( \vartheta_{00}^{2}
(\tau,z) - \vartheta_{01}^{2} (\tau,z) \right) - \vartheta_{10} (2 \tau, 2z)
\vartheta_{10}^{2} (\tau,z) \right] 
= 0 \cr
}}
\eqn\susyqone{\eqalign{
\bra{D3} e^{-TH_{cl}} \ket{\bar{D5; J,M=\half}}  & = \half {q^{h-{1 \over 2}}
\over \eta^{6}(\tau)}  \left[ \vartheta_{10} (2 \tau, 2z) \left(
\vartheta_{00}^{2} (\tau,z) + \vartheta_{01}^{2} (\tau,z) \right) -
\vartheta_{00} (2 \tau, 2z) \vartheta_{10}^{2} (\tau,z) \right] 
 = 0 \cr
}}
We have mutually supersymmetric
branes for any value of the parameter $J$.
Note that the second brane is an anti-brane.  This nomenclature is based on the
semiclassical notion of the flux measured in the weak coupling region. The
two branes above have an opposite sign for the flux (as we will explain in more
detail later on). On the open string side though,
the GSO projection is the same -- so that the two branes are mutually
supersymmetric.
We present a low
energy expansion of these partition sums
 in appendix B.

\ndt {\it Remarks:}
\item{1.}  All the modes is these expansions are localized because the $\ket{D3}$ brane is. 
\item{2.} For $J<0$, both the 
amplitudes $\bra{D3} e^{-TH_{cl}} \ket{D5; J, M=0}$ and $\bra{D3} e^{-TH_{cl}} \ket{\bar{D5; J, M=\half}}$ have a non-unitary spectrum. 
\item{3.} For $0\le J< \half$, the amplitude $\bra{D3} e^{-TH_{cl}}
\ket{\bar{D5; J, M=\half}}$ has a {\it tachyonic spectrum}. The degeneracy
between bosons and fermions then implies that the fermionic spectrum is non-unitary.
\item{4.} For $0<J<\half$, the amplitude $\bra{D3} e^{-TH_{cl}} \ket{D5; J, M=0}$ has massive modes. 

\ndt We are then left with two extended branes with massless modes in their overlap with  $\ket{D3}$,   
the analysis of which we turn to next.

\subsec{Summary of branes which realize interesting gauge theories}
We have the following list of interesting branes which preserve the same
$\CN=1$ supersymmetry in $d=4$ -- the brane $\ket{D3}$, the brane 
$  \ket{\bar{D5; J=M=\half}}$ and $ \ket{D5; J= M=0}$. The various partition
functions contain a four dimensional space filling piece. The massless
spectrum
 in four dimensions is then controlled by the piece of the partition function arising from the cigar. In this subsection, we present some details of the spectra among these various branes and summarize them at the end. 

The brane
$\ket{D3}$ has only localized modes in its spectrum. The cigar piece of the partition function contains only the identity character $Ch_{Id}$ and the massless fields are a gauge field multiplet in four dimensions \Ashok .

The brane $\ket{D5;J=M=\half}$ has no localized modes in its self-overlap. Its
overlap with the $\ket{D3}$ contains from the cigar piece $Ch_{cont}(j=m=\half)$ whose
 massless spectrum consists of a quark and an anti-quark
multiplet  \FNP.

The states $\ket{D5;J,M=0}$ with ${1 \over 4}<J<\half$  have a self- overlap which is exactly the same as that of the $\ket{D5;J=M=\half}$. For $0<J \le {1 \over 4}$, there is in addition one other mode at the boundary of the continuous
representation $j=\half$.  In the overlap with the  $\ket{D3}$, the cigar part
of the character is $Ch_{cont}(j,m=0)$, which 
has generically only massive four-dimensional modes for $J>0$. 

As $J \to 0$, the modes in the overlap with the $\ket{D3}$ start to become massless. 
The brane corresponding to the boundary state
 $\ket{D5;J= M = 0}$ has in its overlap with $\ket{D3}$ the continuous character 
 $Ch_{cont}(j=m=0)$ which is a sum of $Ch_{cont}(j=m=\half)$ and $Ch_{Id}$. 
 The massless fields in this overlap 
are a quark and anti-quark multiplet
 from the $J=\half$ character as well as another field  from the $Ch_{Id}$. The latter looks like it 
 has the quantum numbers of a gauge field with one color and one flavor index. 
The gauge and global symmetries forbid a minimal coupling to the other fields of the 
gauge theory in consideration. 
This field is then not visible at low energies, one can think of it as a localized massive field. 
In the self overlap of $\ket{D5;J=M=\half}$, we find the character $Ch_{cont}(j=m=0)$ arising from a pole 
on which we expand on below.

 \vskip 0.2cm

\ndt {\it A note on Higgsing and bound states}

Before we finish this section with a recap of the highlights in a table, we make a few comments on the realization of the Higgs mechanism in this theory and the interpretation of the characters appearing in the self overlaps of the extended branes. 

There are some characters which arise at the edge of branch cuts in the spectral densities and some which arise from poles. The modes appearing in the first type of character are not localized, they are part of a continuum of modes in six dimensions. On the other hand, a character arising from a pole in the spectral density can give rise to genuinely localized modes. 
This phenomenon also happens in Liouville theory.\foot{For the same technical
reason of the logarithmic branch cut. As we stress later, it can be thought of
as capturing the topological part of the full theory we are interested in.} Indeed, in 
Liouville theory, the full knowledge of the open string correlators showed us
that this mode appearing at the edge of the continuum is not a genuine
localized mode. We can call such modes a marginally localized mode. 

In the tower of states in the character $Ch_{cont}(j=m=0)$, we are interested  in the massless 
modes of which there are two candidate states $j=m=0$ and $j=m=1/2$ both of which are chiral primaries. 
To understand whether one or both of these modes appear as part of the localized spectrum, we appeal to the map between the topological theory on the $SL(2)_{k=1}/U(1)$ coset and the $c=1$ theory at self-dual radius \MukhiZB\ which can be extended to the D-branes and its open strings \AshokXC . According to this map, both the above states are mapped to some open string operators on the $\sigma =0$ brane of Liouville theory \Teschner . The calculation in Liouville theory shows that there is only one genuinely localized mode \Teschner , which according to the map \AshokXC\ maps to the mode with $j=m=\half$.

This mode  $M^{ij}$
transforms in the adjoint of the flavor group and is colorless, {\it i.e.} it
has the quantum numbers of a meson. Recall that the closed string theory and
our branes conserve in perturbation theory the $U(1)_{R}$ symmetry
corresponding to the rotation of the cigar. The meson has charge
$P_{\theta}=1$ under this symmetry. This is twice the charge that the quarks
carry.\foot{Although the quarks and the mesons appear in the same open string
character, the 
quarks have one of their ends on the localized $\ket{D3}$ brane, while the
mesons have both their ends on the extended brane. One can think of the
difference in charges as being absorbed by an open string
 vertex operator which implements the change in boundary conditions.} 
We shall understand this better in geometric terms
 in section \quartic . 

We shall also see in section \quartic\ that the backreaction onto the
Ramond-Ramond axion which effectively counts the objects charged under
$P_{\theta}$ in the massless spectrum is consistent with the existence of one
massless scalar (not two) on the brane $J=M=0$. This convinces 
us that the other marginally massless mode does not affect the four dimensional physics.

To summarize, the $\ket{D5;J=0^{+}, M = 0}$ brane can be understood 
as being at the origin of the Higgs branch of the gauge theory. Thought of as a sum
of the $\ket{D5;J=M = \half}$ and $\ket{D0}$, there are massless quarks in the spectrum. There is a parameter on the $\ket{D5;J=M = \half}$ which gives mass to the quarks. 
On the other hand, the above way of thinking of this 
brane as a single object shows that on this branch, 
the axial combination of the two $U(1)$'s rotating the two 
branes  independently is broken, there is a corresponding 
massless Nambu-Goldstone boson which can be given an expectation value. 
This meson has the same quantum numbers as the operator $Q \T Q$.

 \vskip 0.2cm

{\centerline{Table 1: {\it Summary of mutually supersymmetric  branes with unitary overlaps}}}
\medskip
\centerline{\vbox{\offinterlineskip
\hrule
\halign{\vrule # & \strut\ \hfil #\ \hfil & \vrule # & \ \hfil # \hfil \ & \vrule # & \ \hfil # \hfil \ & \vrule # & \ \hfil # \hfil  \ & \vrule # \cr
height3pt&\omit&&&&&&&\cr 
&{\bf Brane $\ket{B}$}&& RR Charge &&  Overlap with $\ket{D3}$ &&  Self-overlap & \cr
height3pt&\omit&&&&&&&\cr 
\noalign{\hrule}
height3pt&\omit&&&&&&&\cr 
& $\ket{D3}$ &&  +1 &&  Gauge field $A_{\mu}$ &&  $A_{\mu}$    &\cr
height3pt&\omit&&&&&&&\cr 
\noalign{\hrule}
height3pt&\omit&&&&&&&\cr 
& $\ket{\bar{D5;J=M=\half}}$ && $-\half$ && Quarks $Q,\T Q$ && No localized modes &\cr
height3pt&\omit&&&&&&&\cr 
\noalign{\hrule}
height3pt&\omit&&&&&&&\cr 
&  $\ket{D5;J=M=0}$  && $+\half$  &&   Quarks $q,\T q$ &&
Meson $M$  &\cr
height3pt&\omit&&&&&&&\cr 
}
\hrule
}}

\medskip
\leftskip=35pt
\rightskip=35pt
\baselineskip=12pt
\ndt There are other branes with unitary branes with $0<J <\half $ whose
details are in the above subsection. In the last two columns, we show only the
massless genuinely localized modes. 

\leftskip=0pt
\rightskip=0pt
\baselineskip=18pt

\bigskip

\newsec{Relation to brane set-ups in ten dimensions}
\seclab\tend

We can use our understanding of the closed string parameters to link the
non-critical superstring picture to the more familiar brane set-up in ten
dimensions.
A conventional configuration of branes for the study of the physics of $\CN
=1$ gauge theories is the following (see e.g. \GiveonSR\ for a review):
\eqn\branebox{\eqalign{
{Spacetime}: & \quad 0\; 1\; 2\; 3\; 4\; 5\; 6\; 7\; 8\; 9 . \cr
{\rm NS}5: & \quad 0\; 1\; 2\; 3\; 4\; 5\; {\rm -}\; {\rm -}\; {\rm -}\; {\rm -} . \cr
{\rm NS}5': & \quad 0\; 1\; 2\; 3\; {\rm -}\; {\rm -}\; {\rm -}\; {\rm -}\; 8\; 9 . \cr
{\rm D}4: & \quad 0\; 1\; 2\; 3\; {\rm -}\; {\rm -}\; 6\; {\rm -}\; {\rm -}\; {\rm -} . \cr
{\rm D}6: & \quad 0\; 1\; 2\; 3\; {\rm -}\; {\rm -}\; {\rm -}\; 7\; 8\; 9 . \cr
}}
Let us first consider a bulk theory
 without D-branes. The relative motion of the
NS$5$-branes 
is possible in the non-compact directions $(x^{6},x^{7})$. The motion in these
directions is captured in the non-critical string theory in six dimensions by
the parameters 
$\mu, \bar \mu$ of the $\CN=2$ Liouville theory. The string coupling is set by
the absolute value of $\mu$ which measures the distance between the
NS$5$-branes,
while the 
orientation of their relative position
in the $(x^{6},x^{7})$ plane is set by the phase of $\mu$.
The exchange of the NS$5$-branes is implemented by the map 
 $\mu \to -\mu$. 

We can gather further evidence for this identification after introducing the
D$4$-brane into the set-up.
There are three gauge theory parameters for pure Yang-Mills theory -- 
the gauge coupling and theta angle ${1 \over g_{YM}^{2}} + i \theta$, and the
Fayet-Iliopoulos parameter $r$. These three quantities can be thought of as the values that
closed string fields take on the brane. In the non-critical superstring
theory, the relevant string modes are the complex tachyon $T, \bar T$ and the RR axion $\chi$. 
%
For $\mu = \bar \mu $, the modes $(T + \bar T, \p_{+} \chi +  \p_{-} \chi)$ fall into a $\CN=1$ multiplet of the preserved supersymmetries \sameer . The conserved supersymmetries transform covariantly under a rotation of the cigar, and so do the corresponding combinations of the fields. 

The non-critical superstring theory is obtained from the ten-dimensional
string in a double scaling limit (after a T-duality), in which the mass of the D$4$-brane
stretching between the NS$5$-branes is kept fixed while scaling down the 
length of the D$4$-branes, and the string coupling simultaneously \GiveonPX . The fixed mass of the
D$4$-brane sets the parameter $\mu$, which also sets the gauge coupling. (The
phase of $\mu$ does not enter the physics. Also, 
with no flavours present, the
FI parameter cannot be turned on while preserving supersymmetry. We 
recall that we 
have seen in \Ashok\ that the $\theta$ angle is associated with the zero modes
 of the RR axion field $\chi$.

When we further introduce D$6$-branes, the relative orientation of the
D$6$-branes and the NS5 branes in the $(6,7)$ plane becomes
important. Equivalently, the relative orientation of the D6-branes and the
D4-branes stretching between the NS5-branes is fixed by the requirement of
supersymmetry. 

In the non-critical superstring set-up, we certainly have a supersymmetric
 configuration
when the phase of the D$5$ brane boundary state agrees with that of the D$3$
 brane boundary state (in the sense that the overlap is the supersymmetric
one given in the
 previous section). This corresponds to the D6-branes being orthogonal to the
 D4-branes in the ten-dimensional picture.
We can keep the 
 D6-branes fixed while rotating the NS5-branes (and the D4-brane in between),
by changing the phase of $\mu$ and by shifting the angular brane parameter
$M$ simultaneously, thus breaking 
supersymmetry. 

We can thus identify (while keeping in mind that we agree to rotate the
 parameter $M$ of the D6-branes along with the phase of $\mu$)  that the parameter $\half(\mu + \bar
\mu)$ can be associated to the gauge coupling on the D4-brane, which
is the motion in $x^{6}$ (or rather, the
length of the localized D$4$ branes in the $x^{6}$ direction), while the motion in $x^{7}$ is the
FI parameter which we 
associate with ${1 \over 2 i} (\mu - \bar \mu)$.

\subsec{The branes in  the non-critical string theory}
\ndt 
Before re-interpreting our non-critical brane set-up in ten
 dimensions,
let's turn to the brane addition relations in the six-dimensional 
 non-critical superstring.
 We first remark that on subtracting the equations \susyqzero , \susyqone\ with $J=M=0$  and
$J=M=\half$,
 we recover the partition function of the D$3$-brane:
\eqn\susydzero{
\bra{D3} e^{-TH_{cl}} \ket{D5; J=M=0} - \bra{D3} e^{-TH_{cl}} \ket{\bar{D5; J = M=\half}} = \bra{D3} e^{-TH_{cl}} \ket{D3}.
}
We can also use the addition relation of our previous section
 tensored with the same space filling brane in flat space to get the 
brane addition relation in the six-dimensional non-critical superstring theory:
\eqn\braneaddnsixd{
\ket{D5; J=M=0} = \ket{\bar{D5; J = M=\half}} +  \ket{D3} 
}
We will comment later on the addition relation with the branes of opposite charges.

Moreover, we recall that the first order backreaction onto the cigar was calculated in \Ashok , and it
was 
found that the $\ket{D3}$ sources the RR axion. We will see later in Section \quartic\ 
that the $\ket{D5,J=M=0}$ and $\ket{\bar{D5,J=M=\half}}$ also source the
RR axion with charges, in units of the $\ket{D3}$ brane charge $+\half$ for
the first and $-\half$ for the second brane.

\subsec{The identification}
We are ready to identify the branes in the non-critical superstring
theory with the branes in the original (not yet doubly scaled, and T-dual)
ten-dimensional superstring set-up.
For definiteness, let's say the  NS$5$ is to
the left of NS$5'$ in the $x^{6}$ direction. Then we have the following
identifications:
\item{1.} The $\ket{D3}$ maps onto \FNP\ a D4 brane starting on $NS5$ and ending on
$NS5'$. We assign to it a charge $+1$.  (This fixes the orientation of the brane.)
\item{2.} The brane $\ket{\bar{D5, J=M = \half}}$ maps to \FNP\ a  $D6$ brane to the
 left of the $NS5$ with a $D4$ brane starting on it and ending on $NS5$.
 This has charge $- \half$.
\item{3.} $\ket{D5, J=M=0}$ is a $D6$ brane to the left of the $NS5$ with
a $D4$ brane starting on it and ending on $NS5'$.

\ndt The first identification is clear. The brane is the only localized D-brane consistent with unitarity and
supersymmetry which carries the right spectrum, namely a $\CN =1 $ vectormultiplet.
For the second and third identifications, the arguments are the
following. Firstly, these are extended branes,
reaching asymptotic infinity. In the picture T-dual to the cigar, they reach
asymptotic infinity as a particular line in the $(x^6,x^7)$ plane. This agrees
with the asymptotic form of the D$6$-branes in this plane. Secondly, we fixed
the charge of the D$4$-brane before, and the charges listed above are then
found by simple computation. These charges agree with those of the D$6$ branes.
The third argument is that the spectra on these branes agree precisely in the
non-critical and in the ten-dimensional picture. Moreover, a fourth argument
is that indeed, these branes are consistent with the first identification we
made, in that these three classes of branes do satisfy the addition relation
$     \ket{D5, J=M=0} =       \ket{\bar{D5, J=M = \half}} + \ket{D3} $.  The charges add appropriately
$-\half + 1 =  \half$.

Thus, we are now equipped to study the movements of the brane in the ten-dimensional brane
set-up, to translate these into the non-critical superstring theory, and 
 to analyze the effect on the gauge theories living on the branes in terms of
their exact boundary state description.

\newsec{Electric-Magnetic duality in the gauge theory}
\seclab\duality

In this section we discuss electric-magnetic duality  in
$\CN=1$ supersymmetric quantum chromodynamics, within the framework of
non-critical
superstring theory. We have seen how the
$\ket{\bar{D5;J=M=\half}}$ brane introduces  supersymmetric quarks to the
gauge theory on the $\ket{D3}$ brane.
We assumed that $\mu$ is positive in our identification of
these boundary states as positive tension branes.
We also saw that for $\mu>0$, the brane $\ket{D5;J=M=0} $ introduces quarks as well as mesons to the gauge
theory on the $\ket{D3} $ brane.

Although we will not further need the identification of these branes in
ten-dimensions, it may be useful for the reader to keep in mind that we
 identified these boundary states
in the ten-dimensional set-up, under the assumption that the NS$5$-brane is to
the left of the NS$5$' brane. Namely, we identified these boundary states as
corresponding to certain branes in the electric picture. The fact that the
NS$5$-brane is to the left of the NS$5$' brane is equivalent to restricting to
positive values of $\mu$. We want to study now what happens to the
configuration of branes as we go to negative values of $\mu$ (purely within
the exact description of the branes in the non-critical superstring theory).

We start with an electric configuration of $N_{f}$ electric flavor
 branes and $N_{c}<N_{f}$ color branes, and perform the operation $\mu \to
 -\mu$ 
on the system. Using the mapping of boundary states described in equation
\muBtominus , we obtain:
\eqn\electomagbrane{\eqalign{
\ket{D3;\mu} & \to  -\ket{D3;-\mu}; \cr
\ket{\bar{D5;J=M=\half; \mu}} & \to  \ket{D5;J=M=\half; -\mu} \cr
& = \ket{\bar{D5;J=M=0;-\mu}}  - \ket{\bar{D3; -\mu}}. \cr
}}

Now, for negative $\mu$, which we associated to the NS$5$-branes being to the
right of the NS$5$' branes in the $x^6$ direction, we will identify these
boundary states differently with branes in ten dimensions. In particular, we
will know refer to these branes as being magnetic branes.

The final configuration is then $N_{f}$ magnetic branes
 $\ket{\bar{D5;J=M=0;-\mu}}$, $N_{f}$ color anti-branes $(-\ket{\bar{D3;-\mu}})$
 and $N_{c}$ color branes $(-\ket{D3;-\mu})$  (Remember that both of these localized
 branes are well-defined positive-tension objects). Considering  that the
branes and anti-branes 
 annihilate each other and that the left-over purely closed string
 configuration decouples from the gauge theory physics, we find \SenNF:
\eqn\annihilate{
\ket{\bar{D3}} + \ket{D3} \to {\rm closed}\;  {\rm string} \; {\rm vacuum} \;
 {\rm +} \; {\rm closed} \; {\rm  string} \; {\rm decay } \; {\rm products},
}
and we are left with a mutually supersymmetric system of $N_{f}$ magnetic
 anti-branes $\ket{\bar{D5;J=M=0}}$, and $N_{f}-N_{c}$ color anti-branes
 $-\ket{\bar{D3}}$.  Thus, the gauge theory of the dual configuration of
 branes 
 at
negative values of $\mu$ is the Seiberg dual.
Note how the various reversals of sign in \muBtominus\ naturally lead to a reversal of the addition
 relation in the dual configuration, and a set of final states consistent with charge conservation.

Note that for $N_f<N_c$, we cannot condense the open string tachyon fully by this process. In this case, we break supersymmetry. 
In fact, we could describe the above movement of branes  as a function of $\mu$ in the
complex plane. For $N_f>N_c$, the movement can be performed while preserving
supersymmetry along the whole path. While changing the phase of $\mu$, we would rotate the
parameter $M$ of the extended branes as well, keeping the extended branes
fixed at infinity. As we rotate, we need to recombine the extended branes with
the localized branes, turning on a vev for the strings stretching between
$N_c$ of the
localized and the extended branes. It should be possible to
show in detail that these vevs can be turned on consistent with supersymmetry
only when $N_f>N_c$. This follows from the effective action, but it is
feasible as well
to show this explictly using the full boundary state. 
When $N_c < N_f$, we will not find a sufficient number of these open string modes
to preserve supersymmetry while rotating $\mu$.

\subsec{A note on the deformations of the theory}

In the electric and magnetic versions of the gauge theory, it is well-known that the deformations in one theory map to expectation values in the other. In the simplest case, the mass parameter for the quarks in the electric theory has the same quantum numbers as the expectation values of the meson in the magnetic dual. This relation is realized in the open string theory on the branes in a rather interesting manner. 

In the electric description of the theory, the parameter $J$ controls the mass of the quarks for $M=\half$. When $J \to \half$, the quarks become massless and can condense. The mass deformation is described by the zero mode of a field moving on the extended brane. The vertex operator for this field  is described  by the non-normalizable mode corresponding to $h=Q/2=1/2$ occuring in the character $Ch_{cont}(j=m=0)$.
In the dual description, the vertex operator for the open string field corresponding to the dual localized meson 
is almost the same as above with the only difference that the $h=Q/2=1/2$ mode on the cigar takes on the normalizable branch.

\newsec{Tachyon condensation and boundary RG flow}
\seclab\bdryRG

\ndt We have understood the
brane addition relation for a closed string theory $\mu$ 
\eqn\addnelecmag{\eqalign{
  \ket{D5;J=M=0; \mu} & = \ket{\bar{D5;J=M=\half;\mu}}  + \ket{D3; \mu}. \cr
}}
where all the objects in the above equations have positive tension.
We would like to analyze further the analogous relation for mutually
non-supersymmetric
branes in a given closed string theory at a fixed value of the parameter
$\mu$. The first of the above equations \addnelecmag\ implies 
\eqn\branebaraddn{\eqalign{
\ket{D5; J=M=0} + \ket{\bar{D3}} & = \ket{\bar{D5; J= M=\half}}  + \ket{D3} + \ket{\bar{D3}}. \cr
& \longrightarrow \ket{\bar{D5; J= M=\half}}.\cr
}}
where, in the first line, the right hand side manifestly does not preserve any
supersymmetry of the bulk theory. It contains a non-supersymmetric brane without RR charge
(which can decay to the closed string vacuum). 

The general arguments of decay of a brane and its anti-brane to the closed string vacuum are given in the context of open string field theory, and it is understood explicitly as a time-dependent process \SenNF. We wish to argue now that we can actually understand the above equation from the left hand side directly to the second line as a  boundary renormalization group flow on the $J=M=0$ brane. 
We first argue this by relating the set-up to a similar
configuration in bosonic Liouville theory, where the worldsheet boundary RG
flow has been well-understood. Then, we discuss how this could be related to a
boundary RG flow within the cigar boundary conformal field theories.

Firstly, let's discuss how to link up the relation between non-supersymmetric
 boundary states with a boundary renormalization group flow in bosonic
 Liouville theory. By
 appropriately twisting the $\CN=2$ Liouville worldsheet theory, we can focus on the
 topological subsector of our closed string background which is described by a bosonic 
 string theory with a $c=1$ boson at self dual radius coupled to Liouville theory \refs{\MukhiZB, \NakamuraSM }. 
 In \AshokXC , it was
 shown how to extend this map to the open string sector, mapping the boundary
 states and the boundary two-point functions. 
 
 The topological subsector of the
 BPS branes we are studying is thus described by the  branes in Liouville
 theory. The $\ket{D3}$ branes map to the ZZ branes and the extended branes
 $\ket{\bar{D5;J=M=\half}}$ and $\ket{D5;J=M=0}$ map to the FZZT branes
 labelled by $\s = 1$ and $\s = 0$ of \Teschner\ respectively. The addition
 relation we describe above, restricted to the topological subsector  simply
 maps to the addition relation of \refs{ \MartinecKA , \Teschner }
 under the twist. The
 spectrum on the $\s=0$ brane contained in addition to the continuous 
 modes, a localized mode. In the physical non-critical superstring
 theory, this mode  lifts to an infinite set of
 open string modes summarized by the character $Ch(J=M=0)$ which gives rise to the meson multiplet.

We can now try to understand the open string tachyon condensation describing
the process in \branebaraddn\ in this subsector. The  paper \Teschner\  
described how to understand the loss of the localized brane as a boundary RG
flow. The RG flow is seeded by the highly relevant (dimension zero) operator 
which was the new mode  generated on the $\s=0$ brane. In our case, the
operator to which this is lifted is not present in the spectrum of the BPS
branes, due to the GSO projection. However, the spectrum of open strings
between the non-mutually supersymmetric branes $\ket{D5; J=M=0}$ and $\ket{\bar{D3}}$, has the
opposite GSO projection and does contain the relevant boundary perturbation as its lowest mode, the tachyonic
dimension zero operator $B_{0}$. In the topological subsector of the 
Yang-Mills theory we therefore understand the non-supersymmetric 
brane addition relation \branebaraddn . In terms of the pictures
in the previous section, we 
now have the brane folding back on itself and annihilating a little piece of itself.

To carry this over to the full theory, we would need to understand the structure of
the three point functions in the boundary $\CN=2$ Liouville theory, and the
renormalization 
group flows between different boundary conformal field theories. However, we
can already abstract lessons from the bosonic Liouville theory example and
list the properties which will ensure that the RG flow seeded by
the above tachyon proceeds
according to our expectations:
\item{1.} The bulk-boundary correlators for the extended branes  are analytic in
the boundary parameters $J_i$.
\item{2.} The dimension zero boundary operator $B_0$ which is localized  on the
$J=0$ brane will act as a projection operator in the boundary Hilbert space.
\item{3.} The boundary operator $B_0$, when inserted in correlators involving the
boundary state $J=0$ will act as a projector onto the localized brane.
\item{4.} There is an exact boundary renormalization group flow from the boundary
state
$J=0$ to the $J=1/2$ brane, under perturbation by $B_0$.
\item{5.} Thus, the boundary 
RG flow removes the part of the boundary state $J=0$
that is picked up under monodromy, under
perturbation by the localized mode.

\ndt These properties form an important ingredient in a microscopic understanding
of Seiberg duality and it would be interesting 
to demonstrate them beyond our
analysis in the topologically twisted sector (using the results of \Teschner ).

\newsec{What {\it is} the theory on the branes? -- Global symmetries and RG flows}
\seclab\quartic

So far, we have argued that the gauge theories realized as low energy limits
of the two brane configurations belong to the same moduli space. 
To really argue  for a full IR equivalence as in \SeibergPQ\ between the two 
descriptions, one must show that any
open string process that contributes in the extreme IR is 
independent of  $sign(\mu)$.\foot{
This is certainly true for some simple open string processes like those involving only gauge fields
 and its superpartners, and for the mass terms and the quartic coupling of the
 quarks.}

In this section, we try to understand more precisely the relation between the open
string  theory living on the worldvolume of the D-branes 
and the electric/magnetic descriptions of $\CN=1$ SQCD. To this end, we first present
 a list of the global 
symmetries and charge assignments of the various fields. 
We then compute the backreaction onto the closed string background. 
We match the backreaction onto the RR axion with an anomaly coefficient in the field theory which depends only on the massless spectrum. Then we use the backreaction onto the NSNS background to teach us about the interactions in the theory.

From the analysis of the open string theories, we know that 
the two low energy theories under consideration have the same field content as the electric and magnetic descriptions of SQCD. However in our construction, we do not have another 
parameter to tune the QCD scale relative to the string scale and it is determined dynamically.
We can think of the gauge theory on the branes as being completed by an open string field theory.
More practically, this theory can be defined with a string scale cutoff with natural values at that scale for all allowed interactions.  

Although there are only a few terms in the action (corresponding to the ``pure'' SQCD) which are dimensionally relevant, the running to strong gauge coupling invalidates this analysis based on perturbation theory. 
In fact we
know \LeighEP\ that the quartic operator of quark superfields  
$W_{quart} \sim Q\T Q Q \T Q$ although classically irrelevant actually could gain a large anomalous dimension.
Quantum mechanically, this operator depends 
 crucially on the parameters $N_{f}$ and $N_{c}$. SQCD with this 
quartic coupling of the quarks flows to pure electric SQCD if $N_{f}-2N_{c}>0$, but 
 to pure magnetic SQCD 
if $N_{f}-2N_{c}<0$ \StrasslerQS .

We would like to argue that the theory being described is indeed $\CN=1$ SQCD with a quartic coupling\foot{In this context, see also \CaseroPT\ which describes a very closely related setup in supergravity. One would in fact like to argue that our non-critical setup captures the near-singularity region of the setup of \CaseroPT . We thank C. Nunez for a discussion on this issue.} \LeighEP , and that our analysis leads to an exact statement of duality between two theories described by different values of the coupling.

It is difficult to check these statements directly in the full string theory,
since at present we have little knowledge of the three and four
 point functions of open strings in these backgrounds. However,  we can use
the open-closed string duality to gain some insight in how the different
behaviours of the gauge theory depending on the sign of $N_f-2 N_c$ are coded
in the closed string background. 
The first order backreaction on the cigar background can be calculated
as for the case of pure $\CN=1$ super-Yang-Mills in \Ashok . We merely sketch
the calculation here, since it is very analogous to the detailed discussion
in \Ashok . 

\subsec{Backreaction onto closed string background}

To measure the backreaction, we use a similar contour prescription for the
 integral as in \Ashok\ and one has to basically evaluate the one-point
 function multiplied by the profile of the field at a specific value of the
 momentum $p'$ where the integrand has a pole. Since we have measured already
 the backreaction of the localized brane in \Ashok ,
 the relevant quantity is the ratio of one-point functions\foot{ We write here the ratio of the NS-NS sector wavefunctions, there are related expressions for the RR sector.} 
 for the extended  $\ket{D5, p,Q}$ \contbrane\ and the localized $\ket{D3}$ \idbrane : 
\eqn\oneptratio{\eqalign{
{ \psi_{p,Q}(p', Q'=0) \over \psi_{Id}(p', Q'=0)} & =
{  \cos{4 \pi p p'} \over  \sinh{2 \pi p'} \tanh{\pi p'}}\cr 
{ \psi_{p,Q}(p', Q'=1) \over \psi_{Id}(p', Q'=1)} & = 
{ \cos{4 \pi p p'} e^{- i \pi Q Q'} \over   \sinh{2 \pi p'} \coth{\pi p'} }\cr 
}}
To measure the backreaction for the NSNS tachyon with winding number one, and
 the constant mode of the RR field strength, we need to evaluate the above
 expressions at $(ip'=0,Q'=1)$, and to measure the backreaction onto the metric and
 dilaton, we need to evaluate them at
 $(ip'=\half,Q'=0)$. Plugging in these values, we find that 
all three of the above measurements -- the flux of the RR one-form field
 strength, the backreaction onto the graviton-dilaton and the backreaction
 onto the winding tachyon have the following measurements in the units of the
 D3 brane flux\foot{Of course, finding the values for the first 
two branes is enough to find the third because of the addition relation \braneaddn .} :
\eqn\flux{\eqalign{
\ket{D3}: & \quad +1 \cr
\ket{\bar{D5, J= M=\half}}: & \quad -\half \cr
\ket{D5, J= M=0}: & \quad + \half \cr
}}
For $N_{c}$ color branes and $N_{f}$ flavor electric branes, we find that 
 that the backreaction onto the tachyon winding mode, the dilaton and the RR
 axion flux are all proportional 
 to $(2N_{c}-N_{f})$. Note that the evaluation of the backreaction on the
 winding number one tachyon and the RR scalar is not sensitive to the precise
 value of the brane parameter $p$ -- it only depends on the quantized brane parameter
 $Q$
(in constrast to the backreaction on the metric and dilaton).

The backreaction onto the RR axion implies a non-zero theta angle for the
 gauge theory\foot{As a reminder, this
 factor of two arises because of the change of variables 
 between the asymptotic variables and the smooth tip. 
 A rotation of $2
 \pi$ at infinity corresponds to a rotation by $4 \pi$ near the tip, and the
 two angles measured
 respectively phases in the closed string theory and open string theory on the localized brane.} $\theta_{cig}=2\theta_{YM}$  through the coupling of D-instantons \Ashok .
One can also understand in geometric terms  how the rotation of the cigar
affects the various open string modes. Under such a rotation, the quark which
lives on a single sheet of the D2-brane goes around once on a rotation by $2
\pi$ at infinity while the gluino and meson which live on a double sheeted cover of
 the circle go around twice under the same rotation. 
 
 In the gauge theory, the anomaly in the conservation of the current which
rotates the gluini by one unit and the quarks and the anti-quarks in the same
direction by half a unit is proportional to $(2N_{c} - N_{f})$ [see
e.g. \ShifmanUA ]. It is nice check therefore on the consistency of the whole
set-up, that the backreaction onto the zero mode of the RR axion $\chi  =
(2N_{c} - N_{f}) \theta$ indeed measures the anomaly in the 
conservation of the R current which rotates the cigar direction $\theta$  \Ashok .

 We can also check now that these charge 
 assignments indeed generate anomaly coefficients for the rotation $U(1)_{R}$ 
 consistent with \flux\ in the gauge 
 theories on the corresponding branes. In particular, for the $\ket{D5;J=M=0}$, the 
 potential presence of another localized mode with
 non-zero charge under $P_{\theta}$ would 
 be inconsistent with the above analysis. This ties up the loose end of the
 argument
 in 
section \overlap\ for the absence of the marginally massless mode in the low
 energy
 physics.

\vskip 0.2cm

\ndt {\it Comments:}
\item{1.} Measuring the backreaction onto the axion in the corresponding dual magnetic picture ($\T N_{c} = N_{f} - N_{c}, \T N_{f} = N_{f}$) gives the same answer (considering that the charges of all the branes have changed sign). 
\item{2.} The backreaction can really be trusted far from the tip of the cigar, which corresponds to the UV of the open string theory. We might think of the backreaction of one of the NSNS modes as measuring the running of the gauge coupling ${1/g_{YM}^{2}}$. In the region where we can trust this computation, we deduce that the theory on the D-brane at the corresponding energy scale {\it is not} pure electric SQCD (which has a first order  beta function proportional to $3N_{c} - N_{f}$). 
\item{3.} For $N_{f} = 2N_{c}$, we have an extra conserved $U(1)$.

\subsec{Global Symmetries and charges}

For $N_{f}$ flavor branes there is a global symmetry $SU(N_{f})$ rotating the
brane basis. For generic flavor branes at  $J=\half+iP$, it is clear from our
semiclassical discussion (see section \semiclass\ ) of the branes folding over and ending before the tip that one cannot rotate the two sheets of the brane independently. This is consistent with the low-energy theory having a mass term to the quarks. 
As $P \to 0$, the self-overlap \specdenfin\ changes smoothly and there is no extra massless mode that appears at this point. As we move onto the branch $0 \le J \le \half$, there are extra modes appearing at $J={1 \over 4}$ and $J=0$, but these as we saw are localized and have to do with effects in the four-dimensional theory, the bulk densities $\rho_{i}$ actually do not change at all. We view this as an indicator that there  is no symmetry enhancement happening in the six-dimensional theory.\foot{In particular, we never see a $SU(N_{f}) \times SU(N_{f})$ symmetry.} 

There are other $U(1)$ global symmetries of SQCD which are usually written as
 $U(1)_{B} \times U(1)_{a} \times U(1)_{x}$ \ElitzurHC . The charges under
 $(B,a,x)$ of the quarks, gluini and meson are $(\pm 1,1,0)$, $(0,1,1)$ and
 $(0,0,2)$ respectively. The first one, baryon number is exactly
 conserved. The latter two which are both R symmetries are not exact
 symmetries, only a linear combination $R=(1-{N_{c} \over N_{f}})a + ({N_{c}
 \over N_{f}})x$ is \SeibergPQ . In the brane picture, this is understood as
 the fact that although the rotation of the cigar is broken by the presence of
 the branes, there {\it is} a  conserved charge involving a rotation of the
 brane
 itself with respect to the background \FNP . 

The charges in the brane setup we consider are the geometric rotation
 $P_{\theta}$, and the rotation of the left and right part of the extended
 brane\foot{Although there is no $SU(N_{f})^{2}$ symmetry
 on the branes, there is, in perturbation theory a $U(1)^{2}$ as can be seen
 from the fact that there are open strings carrying half-integer momentum in
 the partition function. This is a  phenomenon related
 to the double covering of the brane and exists even for the $J=\half +iP$ branes.} 
 which we combine into their sum and difference $Q^{\pm}$. The charge 
$Q^{+}$ which
rotates the quarks in opposite ways  is the baryon number $B$. Based on our
analysis of the theta angle above and the geometric argument, we understand
that the rotation of the cigar is generated by the charge
 $P_{\theta} = \half(a+x)$.  It is also easy to deduce $Q^{-} =
\half(x-a)$. The charges under $(B,P_{\theta},Q^{-})$ of the quarks, gluini
and meson are $(\pm 1, \half,-\half)$, $(0,1,0)$ and $(0,1,1)$ which 
indeed lead to the usual $(B,a,x)$ assignment.

With these charge assignments, the superpotential of the type $W \sim M Q \T Q$ is allowed (has charge two) in the magnetic theory, both by the classical $R$-charge $P_{\theta}$ of rotation of the cigar, and by the exact R-charge of \SeibergPQ . The rotation $P_{\theta}$ has a natural interpretation as an $R$-symmetry in the closed string background, since the fermions are naturally anti-periodic around the smooth cigar. 
The above coupling of quarks and mesons has total charge $Q^{-}=0$, but
unlike $P_{\theta}$ there is no natural reason to expect $Q^{-}$ to be an
$R$-symmetry even in the classical theory in our six-dimensional setup.

Our background which has a tachyon winding mode condensate has also a closed
 string excitation $X$ of the tachyon with $P_{\theta}=1$. This would have a
 mass of string scale and naturally couple to the quarks as $W \sim X^{2} + \T
 Q X Q$. This superpotential is not forbidden by the classical $R$-symmetry
 $P_{\theta}$, and is a further piece of evidence for the existence of the
 quartic superpotential (at low energies,
 after integrating out the massive field).

\subsec{Softly broken $\CN=2$ theories?}

\ndt The $\CN=1$ SQCD with a quartic coupling can be thought of as arising from a low
 energy description of the softly broken $\CN=2$ gauge theory  with $N_{c}$
 colors and $N_{f}$ matter (hyper)multiplets after integrating out the massive adjoint field.
This theory  indeed has a beta function proportional to the above coefficient $2N_{c}-N_{f}$, and has a global flavor symmetry $SU(N_{f})$.
This gains further support if one thinks of lifting
our whole configuration to ten or eleven dimensions. The six  dimensional
background can be thought of as arising from taking two parallel NS5-branes in
ten dimensions (or M5-branes in eleven) and rotating them  in two of the
dimensions till they are perpendicular \HoriAB  . A non-zero angle of rotation
corresponds to giving a mass to the matter in the vector multiplet (softly 
breaking the theory) and rotation by $\pi/2$ implies the mass 
is of string scale. 

\newsec{Final remarks}
\seclab\final

In this paper, we analyzed in more detail the D-brane boundary states for theories in four
dimensions with $\CN =1$ supersymmetry and $N_c$ colors and $N_f$ flavors,
and their behavior under bulk and boundary transformations. In particular,
the transformation exchanging NS5-branes in the bulk gave rise to a microscopic
description of Seiberg-duality. We discussed a brane addition relation, and
also argued for the existence of a boundary renormalization group
flow of the cigar (or $N=2$ Liouville) conformal field theory that should
encode the projection of a sum of branes onto one of its terms.

Furthermore, we showed that the closed string backreaction captures the
qualitative difference in the behavior of the gauge theory under RG flow,
depending on the sign of the quantity $N_f-2N_c$ as for theories with a 
superpotential quartic in the quark superfields. We also presented other evidence for the 
presence of this operator -- the (absence of) chiral flavor symmetry,  the 
absence of any classical $U(1)$ symmetry forbidding such an interaction, and the enhancement 
of symmetry at $N_{f}=2N_{c}$.

We would like to make a remark about the ``s-rule'' in these systems. In the ten-dimensional setup, this would say that there can be at most one supersymmetric $D4$ brane connecting the $D6$ and the $NS5$ branes. In our configuration, we never see explicitly a free localized $\ket{D3}$ brane on the corresponding extended brane $\ket{D5;J=M=\half}$ (though it carries the corresponding $\ket{D3}$ charge).  The s-rule is in this  sense trivially realized for this object. 
What is interesting however is that the other extended brane $\ket{D5;J=M=0}$ which already carries a  
localized $\ket{D3}$, does not admit\foot{Recall that the spectrum on the  branes with $M=\half$ becomes non-unitary at $J<0$.}  any more. This suggests a corresponding s-rule for the $D6-NS5'$ system. This is another pointer towards an allowed quartic quark coupling. 
Indeed, there is such an s-rule for the  $D6-NS5'$ system with a rotated $NS5'$ which realizes for a finite mass adjoint scale coupled to the quarks \BarbonZU . 


\ndt Finally, some open problems include:
\item{1.} The full proof of the $\CN=2$ Liouville boundary
RG flow and the associated properties of the boundary conformal field theory.
\item{2.} The extension of our analysis to include orientifold planes, and to
realize Seiberg duality in
$Sp$ and $SO$ gauge theories microscopically.
\item{3.} The extension to gauge theories in other dimensions (e.g. three-dimensional 
gauge theories on a $D2+D4$-brane system in type IIA non-critical string theory).
\item{4.} The precise identification of the open string marginal operators
involved in the rotation of the parameter $\mu$, consistently with
supersymmetry. 
\item{5.} Understanding the closed string background after the full backreaction of the color branes, and the special nature of the theory at $N_{f}=2N_{c}$.

\vskip 1 cm

\newsec{Acknowledgments}
We would like to thank Sujay Ashok, Amihay Hanany, Carlos N\'u\~nez and Angel Paredes for discussions. S.M. would like to thank LPT-ENS for hospitality 
while part of this work was being carried out. This work is partially supported by the RTN European program: MRTN-CT2004-503369.

\vskip 1 cm

\appendix{A}{Self-overlap of extended branes}

\ndt We look for poles and the corresponding residues of the spectral density functions defined by \specdenfin\ in the region where we deform the contour. As explained in \TeschnerMD , the contour has to be shifted from $\max(p_{1}, p_{2})$ on the imaginary axis to zero. We look for poles in that range : $0 \le ip \le J$ where $0 \le J \le \half$. To count the number of such poles, the following facts will be useful:
\item{1.} If a function behaves near $x_{0}$ as $f(x) \sim (x-x_{0})^{a}$, then $\p_{x} \log f(x) \sim {a \over (x-x_{0})}$ near $x = x_{0}$. So, the net order of the zero of $f(x)$ (zeros -- poles) is the residue of the function $\p_{x} \log f(x)$.
\item{2.} The function $\CS_{1}(x)$ has simple poles at $x = 2 + m+n$ and simple zeros at $x= -m-n$, with $m,n \in \IZ_{\ge 0}$.

\ndt Putting $p_{1}=p_{2}$ in \specdenfin , we get 
\eqn\isthereapole{\eqalign{
\rho_{1} & = {i \over 2\pi} \p_{p}  \log {\CS_{1} (\half + i{p  })^{2} \CS_{1}(\half + i{2p_{1} }  + i{p })  \CS_{1}(\half - i{2p_{1} }  + i{p })  \over  \CS_{1} (\half - i{p })^{2} \CS_{1}(\half + i{2p_{1} }  - i{p })  \CS_{1}(\half - i{2p_{1} }  - i{p })  } ; \cr
\rho_{2} & = {i \over 2\pi} \p_{p}  \log \CS_{1} (1+ i{p })^{2} \CS_{1}(1 + i{2p_{1} }  + i{p })  \CS_{1}(1- i{2p_{1} }  + i{p }) . \cr
}}
and we 
look for poles in these functions between $p=p_{1}$ and $p=0$, with $0 \le
ip_{1} \le \half$. (See the bulk of the paper.)

\appendix{B}{Expansion of some brane overlaps}

We present here the low energy expansion of the overlaps between the branes $\ket{D5;J,M}$ with $\ket{D3}$. Let us write a generic partition sum as $Z = Z^{NS} +  Z^{\T{NS}} + Z^{R} + Z^{\T R}$. The low energy expansions of these functions are (with $h=-J(J-1)+M^{2}$):
\eqn\pfnexpanqzero{\eqalign{
& \quad {\bf M=0} \cr
Z^{NS} & = q^{h-\half} + 4 q^{h} y + .. \cr
Z^{\T{NS}} & =  - q^{h-\half} + 4 q^{h} y + .. \cr
Z^{R} & =   -4 q^{h} y + .. \cr
Z^{\T{R}} & =  0 \cr
}}
\eqn\pfnexpanqone{\eqalign{
& \qquad \quad {\bf M=\half} \cr
Z^{NS} & = 2 q^{h-\half} y - 8 q^{h} y^{2} + 4 q^{h+ \half} y^{3} + ... \cr
Z^{\T{NS}} & =  2 q^{h-\half}y + 8 q^{h} y^{2} + 4 q^{h+ \half} y^{3}+ ... \cr
Z^{R} & =   -2 q^{h-\half} y + 0 q^{h} y^{2} - 4 q^{h+ \half} y^{3} + ... \cr
Z^{\T{R}} & =  0. \cr
}}

\listrefs

\end